# Quantifying attrition in science: a cohort-based, longitudinal study of scientists in 38 OECD countries

Marek Kwiek[1,2] · Lukasz Szymula[3,4]

## Abstract

In this paper, we explore how members of the scientific community leave academic science and how attrition (defined as ceasing to publish) differs across genders, academic disciplines, and over time. Our approach is cohort-based and longitudinal: We track individual male and female scientists over time and quantify the phenomenon traditionally referred to as "leaving science." Using publication metadata from Scopus—a global bibliometric database of publications and citations—we follow the details of the publishing careers of scientists from 38 OECD countries who started publishing in 2000 ($N = 142,776$) and 2010 ($N = 232,843$). Our study is restricted to 16 STEMM disciplines (science, technology, engineering, mathematics, and medicine), and we track the individual scholarly output of the two cohorts until 2022. We use survival analysis to compare attrition of men and women scientists. With more women in science and more women within cohorts, attrition is becoming ever less gendered. In addition to the combined aggregated changes at the level of all STEMM disciplines, widely nuanced changes were found to occur at the discipline level and over time. Attrition in science means different things for men versus women depending on the discipline; moreover, it means different things for scientists from different cohorts entering the scientific workforce. Finally, global bibliometric datasets were tested in the current study, opening new opportunities to explore gender and disciplinary differences in attrition.

✉ Marek Kwiek
kwiekm@amu.edu.pl

[1] Center for Public Policy Studies (CPPS), Adam Mickiewicz University of Poznan, Poznan, Poland

[2] German Center for Higher Education Research and Science Studies (DZHW), Berlin, Germany

[3] Faculty of Mathematics and Computer Science, Adam Mickiewicz University of Poznan, Poznan, Poland

[4] Department of Computer Science, University of Colorado Boulder, Boulder, CO 80309, USA

## Introduction

In the present research, we explore how members of the scientific community leave academic science (in our conceptualization, how they cease publishing in academic journals) and how attrition differs across genders, academic disciplines, and over time. Our approach is cohort-based (Glenn, 2005) and longitudinal (Menard, 2002; Singer & Willett, 2003): We track male and female scientists over time and quantify the phenomenon traditionally referred to as "leaving science" (Geuna & Shibayama, 2015; Preston, 2004; White-Lewis et al., 2023; Zhou & Volkwein, 2004). Our focus is on leaving science, which can be seen as ceasing scholarly publishing because large-scale longitudinal data on leaving academic employment—which would be more adequate—are not currently available at a global level.

Using bibliometric publication metadata from Scopus—a global bibliometric database of publications and citations—we follow the details of the publishing careers of 142,776 male and female scientists from 38 OECD countries who began publishing in 2000 and 232,843 scientists who started publishing in 2010 (termed the 2000 and 2010 cohorts). Our study is restricted to 16 STEMM disciplines (science, technology, engineering, mathematics, and medicine), and we track the individual scholarly output of the two cohorts until 2022.

Bibliometric metadata used in this study are a perfect example of the digital traces left in publications used to study scientists who have traditionally been explored through surveys, interviews, and administrative and census data. Digitalized scholarly databases with bibliometric information are a new source for studying scientists as a population, even though they first need to be repurposed to focus on individual scientists rather than individual publications. This allows for the exploration of questions about science and scientists at an unprecedented level of detail (Kashyap et al., 2023; Liu et al., 2023; Wang & Barabási, 2021). Individuals can be studied according to age, seniority, gender, discipline, and institutional type—and most importantly, for the present study, scientists can be tracked over time. Moreover, cross-sectional studies can be complemented by longitudinal studies, in which individual academic careers are tracked for years and decades. Using massive datasets, academic careers have been examined globally and nationally (e.g., Ioannidis et al., 2014; King et al., 2017; Larivière et al., 2013; Milojevic et al., 2018; Nielsen & Andersen, 2021; Nygaard et al., 2022; Kwiek & Szymula, 2023; Spoon et al., 2023).

Traditionally, women tend to leave academia earlier than men do. Moreover, they tend to do so in greater proportions than men, as the traditional narrative on gender differences in attrition in higher education suggests (Alper, 1993; Blickenstaff, 2005; Deutsch & Yao, 2014; Goulden et al., 2011; Preston, 2004; Shaw & Stanton, 2012). In our examination of attrition, we move beyond traditional cross-sectional approaches (single or repeated snapshots in time) and instead use large-scale longitudinal data at the level of individual scientists from 38 OECD countries. Furthermore, we test new possibilities opened up by global bibliometric datasets for the large-scale quantification of attrition in scientific (publishing) careers.

## Theoretical contexts

### Male scientists, female scientists, and attrition in science

Both young male and female scientists face various barriers to entering and pursuing academic careers (Preston, 2004; Wohrer, 2014); however, attrition has traditionally been examined as a specifically female-dominated phenomenon: Women have been reported as facing "chilly" workplace cultures, difficulties in maintaining work–life balance, and difficultly surviving motherhood periods while working in academia (Cornelius et al., 1988; Goulden et al., 2011; Maranto & Griffin, 2011; White-Lewis et al., 2023; Wolfinger et al., 2008; Britton, 2017). The major dimensions of these barriers are academic promotions, research productivity, scholarly impact, access to research grants, awards, and recognition of scholarly achievements: Women are underrepresented in senior academic positions; have a lower likelihood of collaborating internationally in research, publishing in high-impact journals, and being highly cited; and have a greater likelihood of longer career breaks and grant rejections (Hammarfelt, 2017; Kwiek & Roszka, 2021a, 2021b; Shibayama & Baba, 2015; Sugimoto & Larivière, 2023; Tang & Horta, 2023). In the majority of STEMM disciplines, women enter predominantly male-dominated environments in which "old boy" networks operate and where a "chilly climate" dominates (Santos et al., 2020).

Although both men and women leave academic science in high proportions—approximately one-third of all publishing scientists leave science within the first 5 years and about a half within a decade (as we later show for the 2000 cohort)—attrition is believed to be greater for women than for men (Kaminski & Geisler, 2012; Preston, 2004). The "leaky pipeline" and "chilly climate" hypotheses explain this difference for STEM disciplines: There is a loss of talent at every stage of the academic career pipeline because of systemic barriers for women (Blickenstaff, 2005; Goulden et al., 2011; Shaw & Stanton, 2012; Wolfinger et al., 2008), and a hostile or unwelcoming work environment can discourage women from pursuing their careers (Cornelius et al., 1988; Spoon et al., 2023).

According to the leaky pipeline model, individuals either progress through a series of academic stages or leave academia altogether. The "chilly climate" for female faculty in STEM disciplines—the "perception of exclusion" and a sense of "not belonging"—is grounded in relational demography (low percentages of women within selected disciplines), but there are also other individual antecedents; these include the percentage of women in a particular department, procedural fairness, and gender equity (Maranto & Griffin, 2011, p. 143–146). Importantly, from a policy perspective, the "leaky pipeline" has made prolonged and significant efforts to promote gender equity on campuses in the US and across Europe.

The literature on women in science reflects two major perspectives (Branch, 2016; Fox, 2020). The science "pipeline" perspective is about individuals, and the science "pathways" perspective is about the organizational structures in which individuals are set. The science pathways perspective also addresses dynamic processes and patterns that are not necessarily orderly progressions from one stage to another. Importantly, the structures are not fixed forever; they are changeable. The two perspectives have different practical consequences and may lead to different practical steps, possibilities for intervention, and change (Fox & Kline, 2016, p. 63).

The "pipeline" metaphor highlights straight links between educational stages and between education and occupational outcomes (Xie & Shauman, 2003). The focus of the literature based on the pipeline metaphor is on individual women and their alleged shortcomings (compared with men); if corrected, these shortcomings will decrease the leaking of women from the scientific pipeline. Attention is given to individual women instead of the structural conditions of science, in which women are embedded as students, doctoral students, and finally scientists. The pipeline metaphor, being primarily about "keeping women in the pipeline" (Fox & Kline, 2016, p. 57), assumes a passive flow of women (and men) from one academic career stage to the next. Persistence in the pipeline depends on increasing the supply of women early on; more women, such as students, doctoral students, and early-career researchers, are viewed as a remedy. The metaphor assumes equitable conditions for women along the way, and it does not address the questions of who leaks and why (Branch, 2016, p. 7).

The pipeline metaphor dismisses signals about who naturally belongs to science and who is naturally excluded from it, who is the stereotypical ideal scientist (in mathematics, computing, engineering, etc.); it dismisses gendered obstacles and disregards "persistent exclusionary messages" directed at women (Branch & Alegria, 2016, p. 8). The pipeline conceptual perspective leads to the conclusion that the underrepresentation of women in science is attributable to women's relatively low supply and relatively higher rates of attrition from the science pipeline; as a result, the practical answer to this perspective is to block leakage (Xie & Shauman, 2003, p. 7) and to increase the supply of women (Fox & Kline, 2016, p. 63).

The "pathways" metaphor, in contrast, emphasizes progression in science that is neither direct nor simple and focuses on the organizational culture in which scientists are embedded and on the settings in which they work. What matters for progression in science is more than individual traits; it is the actual organizational setting in which women operate in science on a daily basis. Organizational culture involves shared values, beliefs, and behaviors within organizations; a dominating culture defines what is more (and less) valued in organizations and promotes "ways of doing business."

Solutions within the first perspective include individuals and how to make them fit better with the (traditionally masculine) idea of doing science; solutions within the second perspective include institutional changes and how to make departments more responsive to inequalities in gender, rank, and rewards (Branch, 2016; Fox & Kline, 2016; Xie & Shauman, 2003). The pathway model emphasizes dynamic processes in settings in which women are educated and employed and complex outcomes, not necessarily orderly progress from one stage to another (Fox, 2020). The pathway model allows for nonparticipation in science at any stage, late entry into science professions, and the positioning of science careers in the context of other life course events (such as family formation) (Xie & Shauman, 2003).

Finally, the "chilly climate" hypothesis needs to be distinguished from the wider idea of different "climates" in academic science departments. The idea of departmental climates comes from the realization that scientists have individual characteristics but do not exist in a social vacuum; organizational conditions in academic science departments are important (Fox, 2010). Scientific work relies upon the cooperation of others and requires human and material resources (Fox & Mohapatra, 2007). To understand the status of women in academic science, the features of the organizations in which they work need to be examined. "Scientific work is fundamentally social and organizational" (Fox, 2010, p. 1000). These organizational factors as frequency of discussion about research, interactions, and exchanges about research, space available for research, recognition and rewards received in home units, and work–family interference shape participation and performance in science. What influences science is the "personality," "character," or "climate" of the home units in which science is performed and how the home unit culture is perceived. Women in departments may be more or less included in their social milieu and may remain inside or outside heated discussions and social networks (Fox, 2010). Departmental climate encompasses perceptions about values and practices within an organizational setting and can be measured through surveys (e.g., climate can be measured along bipolar dimensions, such as unfair–fair, stressful–unstressful, or noninclusive–inclusive; see Fox & Mohapatra, 2007, p. 141).

Although, ideally, in academic career studies, individual-level and organizational-level analyses could be combined, and quantitative research results could be examined along with qualitative research results, our variables are relatively limited in scope. Our focus is descriptive, and our data are quantitative and individual in nature. We have no access to variables and measures related to organizations and their units and no access to individual traits other than those derived from bibliometrics. We also have no access to individuals' perceptions of the way things are around in their departments (climates, cultures, or "ways of doing business" in science at the lowest organizational level). To analyze these dimensions of doing science, a global survey of academics currently under preparation is needed. However, we are able to follow large numbers of scientists for more than two decades and longitudinally analyze gender differences in their publishing careers.

### Global datasets and big data approaches to science and scientists

Attrition in science can be quantified beyond individual institutions and countries, and large-scale datasets can be used for this purpose. Global and longitudinal approaches to academic (publishing) careers have only recently been made possible by increasing access to digital databases with comprehensive information about scientists, their research outputs, and their citation-based impact on global scholarly conversations (Kashyap et al., 2023; Wang & Barabási, 2021). The advent of new digital datasets, access to immense computing power, and a more general turn toward structured big data in social science research have led to a recent explosion of studies about the various aspects of academic careers, with an impressive line of research focused on the differences between men and women in science from various perspectives (e.g., King et al., 2017; Nielsen & Andersen, 2021; Sugimoto & Larivière, 2023).

Large datasets provide the unique capacity to test traditional beliefs and conceptual frameworks about science and scientists (Liu et al., 2023). Digital data that trace the entirety of a scientific enterprise can be used to capture "its inner workings at a remarkable level of detail and scale" (Wang & Barabási, 2021, p. 1). It is now possible to systematically explore the publishing career histories of hundreds of thousands of individual scientists and the details of their careers. Today, massive datasets are available at researchers' fingertips, although not without new limitations (e.g., Liu et al., 2023; Sugimoto & Larivière, 2023).

### Leaving science as a scholarly theme

However, the theme of leaving science has not been comprehensively examined either globally or from a multicountry perspective. It has traditionally been explored either through small-scale case study research (mostly survey- and interview-based research) or through multiyear US studies of postsecondary faculty (e.g., Rosser, 2004; Xu, 2008; Zhou & Volkwein, 2004). Most recently, White-Lewis et al. (2023) have examined the actual departure decisions of 2289 US faculty members who left their institutions between 2015 and 2019. Overall, women were found to leave academia at higher rates than men at every career age. Importantly, for women, workplace climate matters more than work–life balance in leaving their academic positions (Spoon et al., 2023). Our findings generally confirm a recent US case study (Spoon et al., 2023) showing that women leave academia overall at higher rates than men at every career stage, especially at lower-prestige institutions. The authors examined a large employment census of tenure track or tenured faculty active within a 10-year period across 111 academic fields (using Academic Analytics data), along with a broad survey about attrition. The authors, using survey data, highlight that, independent of career stage, women are substantially more likely to report feeling pushed out, and men are more likely to report feeling pulled toward an attractive opportunity when they leave. Workplace climate is a major reason that women leave academia (rather than work–life balance)—which, however, cannot be examined without survey or interview data.

Leaving science has been previously studied through concepts such as "faculty departure intentions" (Zhou & Volkwein, 2004), "intentions to leave" (Rosser, 2004), "faculty turnover" (Ehrenberg et al., 1991), and "faculty turnover intentions" (Smart, 1990; Xu, 2008). The majority of attrition studies have focused on a single institution, and its geographical scope has been limited to the United States (exceptions include Milojevic et al., 2018, who studied astronomy, ecology, and robotics globally using bibliometric data from principal journals belonging to these fields).

Previous studies have shown that women's stronger turnover intentions are highly correlated with an academic culture that provides women with fewer advancement opportunities and limited research support relative to men's and women's roles vis-à-vis family responsibilities (Xu, 2008). However, disciplinary variations in faculty turnover matter, as academics in different disciplines exhibit varying attitudinal and behavioral patterns, and there are distinct opportunities offered inside and outside academic environments (Zhou & Volkwein, 2004).

What is important for us in this paper is what Kanter (1977) termed the "proportional scarcity" of women in those disciplines in which women scientists have traditionally occupied token status—where they were either alone or nearly alone in a peer group of men scientists. As we have shown elsewhere in great detail (Kwiek & Szymula, 2023), the disciplines with highly skewed gender ratios are mathematics, physics, astronomy, computer sciences, and engineering, according to the Scopus classifications (MATH, PHYS, COMP, and ENG). "Tokens" (i.e., minorities among "dominants," with a ratio of approximately 15:85—which is generally the current proportion of women in these disciplines) are often treated as representatives of their categories and "as symbols rather than individuals" (Kanter, 1977, p. 54). All of their actions are public; they are visible as category members, and their acts (here under performance pressure to publish) tend to have added symbolic consequences.

Academics' perceptions of their work lives have been reported to exert a direct impact on their satisfaction and, subsequently, their intention to leave (Rosser, 2004). Academics are pushed to stay/leave their current institutions by a number of internal forces, which can be classified into three major clusters of factors: organizational characteristics, individual characteristics, and work experiences. These factors influence faculty job satisfaction, which in turn influences intentions to leave. They are also pulled by a number of external factors to leave their institutions.

Internal factors include individual and family characteristics (e.g., gender and family/marital status), organizational characteristics (e.g., wealth or unionization), and work experiences (e.g., workload, productivity, and compensation). The external factors include the following: the external job market, extrinsic rewards, research opportunities, teaching opportunities, and family considerations. Internal factors directly influence job satisfaction and perceptions of the organizational environment, which influence intentions to leave; external factors have also been shown to either strengthen or weaken intentions to leave (Zhou & Volkwein, 2004, p. 144–147). Smart (1990, pp. 406–409) proposed a causal model to assess the predictors of the intention to leave a current institution for another position in either an academic or nonacademic setting. The three major sets of determinants were individual characteristics (e.g., age and working time distribution), contextual variables (e.g., salary, influence, and career satisfaction), and external conditions (e.g., economic and societal conditions).

The decision to make a career move involves a comparison of the expected present values of the pecuniary and nonpecuniary conditions of employment at the current institution and its alternative (Ehrenberg et al., 1991). From an economic perspective, depending on salary structures, different decisions to stay/leave can occur: In institutions with low salary dispersion, the most productive academics may tend to leave (because they feel undercompensated), whereas in institutions with high salary dispersion, less productive academics may tend to leave (because they feel underpaid relative to their colleagues; Ehrenberg et al., 1991, pp. 107–108). Smart (1990) has shown that being male, spending more time on research, and being more productive positively influence the intentions of tenured faculty to leave, whereas salary satisfaction is an influential variable only for nontenured faculty.

The conceptual frameworks employed to study faculty rationales for leaving higher education include "push" and "pull" factors (i.e., features of the current academic workplace and external features that attract faculty outside their institutions; White-Lewis et al., 2023). However, if push factors are minimal, then pull factors are expected to have less weight in departure decisions (White-Lewis et al., 2023). Explanations of quitting science include problems of work–life balance (Rosser, 2004; Smart, 1990), low job security, low salaries (Zhou & Volkwein, 2004), colleagues, and workload concerns (Wohrer, 2014) as well as various types of discrimination in the workplace (Preston, 2004; Smart, 1990) and hostile workplace climates (Cornelius et al., 1988; Spoon et al., 2023).

There are important conceptual differences between "leaving the institution" and "leaving academia," between "intentions to leave" and "actual departure decisions" (White-Lewis et al., 2023), and our conceptualization, in which "leaving science" is examined through the notion of "not publishing" in academic journals any longer. Our focus in the survival analysis on publishing over the years until the event of finally "not publishing" at some point goes beyond institutions and sectors to a more general level (publishers vs. nonpublishers in academic science).

## Data and methods

### Scholarly publishing events and survival analysis

Scientific life can be conceptualized as a sequence of scholarly publishing events, from the first publication event to subsequent publications and, in many cases, to the most recent publication ever, when scientists simply cease publishing. In our study, scientists who published for the first time in 2000 composed the 2000 cohort from which, gradually, year by year, both men and women scientists attrit at varying rates of intensity.

Leaving science is conceptualized in our research as an event and analyzed within what is termed *survival analysis* (Allison, 2014; Mills, 2011). Although the general theme of "leaving science" has been widely explored (Geuna & Shibayama, 2015; Preston, 2004), leaving science as an event marked by ceasing publishing has not been extensively studied using survival analysis (and, to the best of our knowledge, has not been studied before from a large-scale multicountry quantitative perspective).

In survival analysis, questions related to the timing of and the time span leading up to the occurrence of an event are explored (Mills, 2011). An event of interest is the final publication marking the final year of remaining in science; the time span leading up to the event of leaving science is referred to as the *survival time*. A classic statistical technique for survival analysis is the Kaplan–Meier estimate of survival (Mills, 2011). A plot of the Kaplan–Meier estimator is a series of characteristically declining horizontal steps of various heights.

When some of the subjects of a study did not experience the event before the end of the study (as in our case, in which some scientists continued publishing during the study period and did not "leave science" according to our definition of the term), they were termed right-censored observations. For right-censored observations, we have partial information: The event may have occurred sometime after or just before the last year of our study (or will occur), but we do not know exactly in which year.

In our case, authors whose last publication was dated 2019 or later were marked as censored observations (i.e., leaving science in 2020 or some time afterward). To classify an author as leaving science or staying in science, the final publication must be dated to 2018 (marked as leaving science in 2019) or earlier. Uncensored cases represent observations for which we know both the starting year (2000 for the 2000 cohort, 2010 for the 2010 cohort) and the ending year of being in science, which must be 2019 or earlier (determined by the date of the last publication and leaving science the following year; see "Publishing breaks and publishing frequency" in the Electronic Supplementary Material or ESM).

For each year, the initial number of scientists entering the time interval consists of scientists who leave science during this interval and those who stay in science in the following interval. The total probability of survival until a given time interval is calculated by multiplying all probabilities of survival across all time intervals preceding that time (Mills, 2011). The two survival curves—for men and women—can be compared statistically to test whether the difference between survival times for the two groups is statistically significant.

Compared with previous studies, the current research presents a different geographical scale (38 OECD countries combined), thus moving away from single-country institutionally focused case study research designs based on surveys and interviews and toward multicountry research focused on academic disciplines. Moreover, the current study examines cross-disciplinary and gender differences in attrition over two decades via a longitudinal approach to large, nonoverlapping cohorts of scientists. We test the power of structured, reliable, and curated big data (of the bibliometric type—the Scopus dataset).

The present research is also longitudinal in the strict sense of the term; rather than using a cross-sectional (one snapshot) or repeated cross-sectional (several snapshots) approach, cohorts of scientists have been tracked over time for up to two decades (2000–2022) on a yearly basis. Consistent variables and stable classifications are used across national science systems, and across time, a wealth of individual micro-level data is leveraged, and time is regarded as a critical variable.

## Dataset and dataflow: what can we know about individual scientists from publication metadata?

The present study uses publication and citation bibliometric metadata about researchers starting to publish in the Scopus database for the first time in 2000 and 2010 (as well as in the years in between). The full population of researchers from the 2000 to 2010 cohorts was $N = 2,127,803$ (Supplementary Table 1, panel 1). Scopus is a large global abstract and citation database of peer-reviewed literature, and it is particularly suitable for global analyses at the micro-level of individual scientists because it is well organized around Scopus Authors IDs, apart from the focus on publications and their metadata (Baas et al., 2020). The dataflow (Fig. 1) shows two major cohorts: scientists starting to publish in 2000 (left) and scientists starting to publish in 2010 (right). The steps for both cohorts were as follows: to define scientists with at least two articles (or papers in conference proceedings) in their publishing portfolios, to define their country affiliation as an OECD country, to define their gender (binary: male or female), and to define their discipline as STEMM.

We used all 16 STEMM disciplines, as defined by the journal classification system of the Scopus database (all science journal classification, ASJC): AGRI, agricultural and biological sciences; BIO, biochemistry, genetics, and molecular biology; CHEMENG, chemical engineering; CHEM, chemistry; COMP, computer science; EARTH, earth and planetary sciences; ENER energy; ENG, engineering; ENVIR, environmental science; IMMU immunology and microbiology; MATER, materials science; MATH, mathematics; MED Medicine; NEURO neuroscience; PHARM pharmacology, toxicology, and pharmaceutics; and PHYS, physics and astronomy.

The scientists in our sample were assigned to gender, academic discipline, country, and year of exit from academic publishing (rather than exit from science sector employment or higher education employment, Table 1). The determination of other data points for individual observations is discussed in the ESM (the year of the start of the publishing career, and the publication minimum).

First, the validity of gender determination algorithms is of critical importance in any study that contrasts men and women in science. The gender of the authors is not disclosed in peer-reviewed publications (Science-Metrix, 2018). Therefore, using articles' author name information is the only means available in large-scale multicountry studies, such as ours, to compare men and women in science. There are several limitations in inferring gender from bibliometric databases: First names are not always available (sometimes only initials are provided); the proportion of papers for which first names are available varies over time, across fields and subfields, and between countries; and not all first names are gender specific. Scopus is preferred to the Web of Science in our study because the Web of Science cannot be used to produce indicators on gender at the country level before 2008 (Science-Metrix, 2018), and we cover the period 2000–2022.

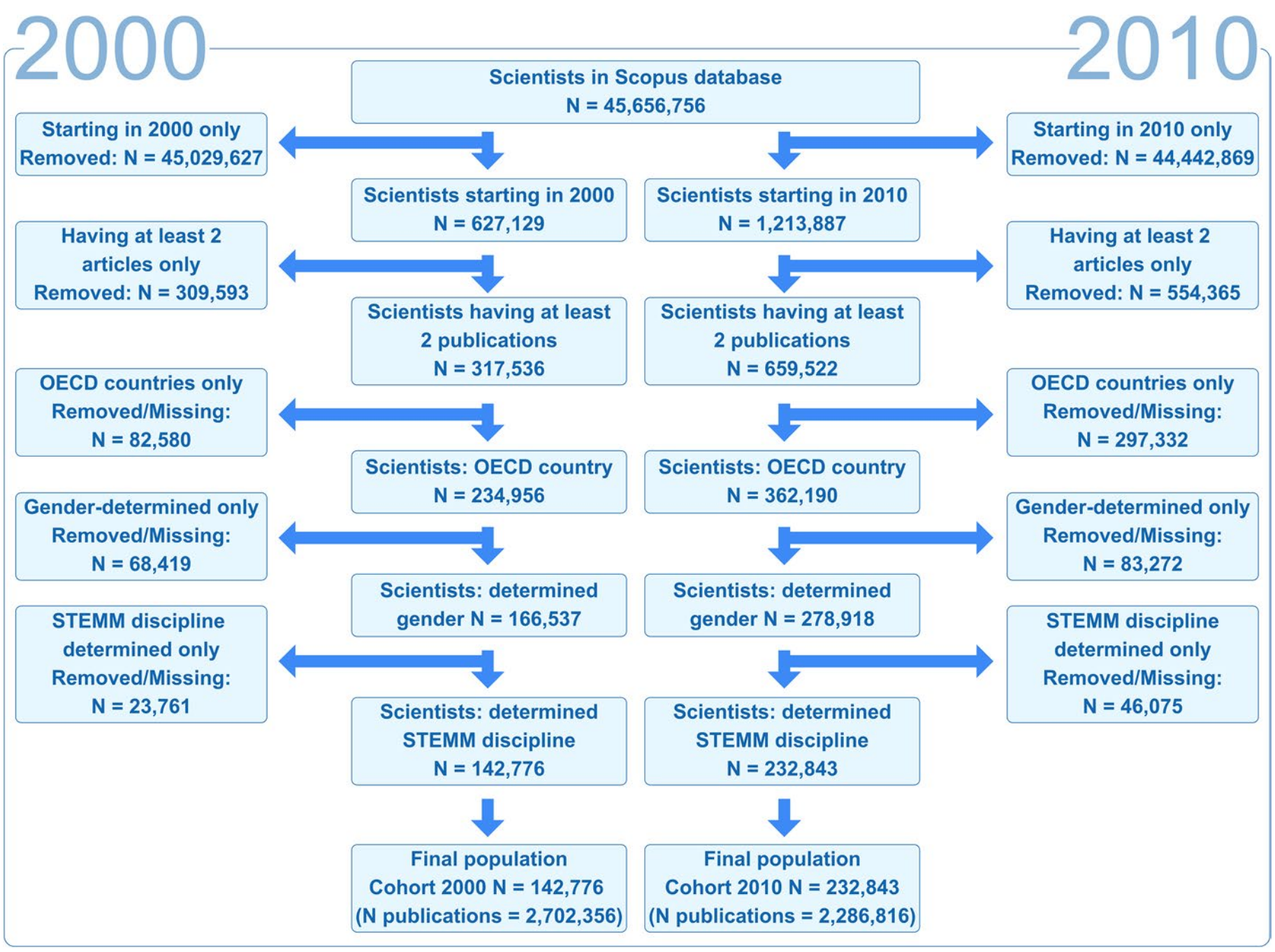

**Fig. 1** Dataflow: subsequent steps to define the 2000 (left) and 2010 (right) cohorts of scientists

In our previous smaller-scale national-level studies (with $N=25{,}463$ scientists, e.g., Kwiek & Roszka, 2021a; Kwiek & Roszka, 2021b; Kwiek & Roszka, 2022), we used gender information provided by an administrative and biographical dataset (governmental data: the national registry of scientists), and in our previous survey-based studies, we used self-declared gender information (Kwiek, 2016; Kwiek, 2018). However, no administrative data at the micro level of individuals are available for the OECD countries studied in the present research. As a result, to determine the scientists' gender in our study, the approach applied in two recent Elsevier reports on women in science (Elsevier, 2018; Elsevier, 2020) was used. The authors' gender inference data were made available to us through the International Center for the Study of Research (ICSR) Lab platform based on a multiyear collaboration agreement.

The dataset we used contained the author's identifier from the Scopus database and two variables determined using the NamSor gender inference tool. NamSor offers a high degree of accuracy (i.e., there are few false positives) and recall (i.e., there are few unknowns) and global coverage. Their validation procedure relies on the use of directories' listing names and geographical locations (Science-Metrix, 2018; NamSor, 2024). The NamSor API (application programming interface) was validated in a report on the development of

**Table 1** Example of the two-cohort database: micro-level data for selected scientists from cohort 2000 (panel 1) and Cohort 2010 (panel 2), $N=375{,}619$

| Scientist ID, the two-cohort database | Gender | Discipline | Country affiliation | Year entering science (year of first publication) | Year leaving science (year of last publication plus 1) |
|---|---|---|---|---|---|
| **Panel 1: scientists from the 2000 cohort ($N=142{,}776$)** | | | | | |
| ID 1 | Female | MED | Spain | 2000 | 2020* |
| ID 2 | Male | COMP | The USA | 2000 | 2004 |
| ID 3 | Female | AGRI | France | 2000 | 2008 |
| ID 4 | Male | PHYS | Japan | 2000 | 2013 |
| ID 5 | Female | CHEM | Denmark | 2000 | 2001 |
| ID 142776 | Male | MED | Germany | 2000 | 2017 |
| **Panel 2: scientists from the 2010 cohort ($N=232{,}843$)** | | | | | |
| ID 142777 | Male | ENER | The UK | 2010 | 2012 |
| ID 142778 | Female | IMMU | Switzerland | 2010 | 2020* |
| ID 142779 | Female | BIO | Belgium | 2010 | 2017 |
| ID 142780 | Male | ENG | Canada | 2010 | 2014 |
| ID 142781 | Male | MED | Italy | 2010 | 2012 |
| ID 375619 | Female | AGRI | Australia | 2010 | 2015 |

*Right-censored observations

bibliometric indicators to measure women's contribution to science using data from the Official Directory of the European Union (recall: 97.8% for women and 98.5% for men; precision: 97.2% for women and 98.8% for men) and using data about the Olympic medalists (1960–2008) (recall: 96.8% for women and 97.2% for men; precision: 94.2% for women and 98.5% for men) (Science-Metrix, 2018).

For the OECD countries studied in our research, NamSor works very well with most names in Western countries, well with Japanese names, and much worse with Korean names (China is not covered, with it not being an OECD member state). However, we did not exclude South Korea because, in the aggregate approach used in the study (all OECD countries examined), there were only 3152 Korean observations in the 2000 cohort and 7703 Korean observations in the 2010 cohort (2.21% and 3.31% of the sample, respectively). Our previous study of the changing demographics of the global scientific workforce revealed that 19.31% of scientists in all STEMM disciplines combined were female scientists in South Korea in 2021 (Kwiek & Szymula, 2023). Unfortunately, NamSor software offers special features for detecting the gender of Chinese and Japanese names, but not for detecting the gender of Korean names.

The NamSor software used for gender detection in our dataset has been positively evaluated in several studies, and recent studies comparing the various commercial gender detection tools have shown that NamSor provides the best results. A systematic comparison of gender detection methods was first presented by Karimi et al. (2016), following several previous major applications of male/female distinction in global studies of academic science (e.g., West et al., 2013; Larivière et al., 2013; Mihaljević-Brandt et al., 2016; Holman et al., 2018). Using a dataset of 7,076 manually labeled names, Santamaria and Mihaljević (2018) studied five commercial name-to-gender inference services and concluded that the Gender API and NamSor were the most accurate tools. The accuracy and misclassifications of the four major gender detection tools were studied, and Gender API and NamSor were found to be the most accurate tools (Sebo, 2021). NamSor was also found to perform well in predicting the country of origin and ethnicity of individuals based on their first and last names, but using the level of continents rather than countries was suggested for research practice (Sebo, 2023).

The gender probability score in NamSor is based on three input parameters: the author's first name, the author's last name, and the author's dominant country from the first year of publication in Scopus ($N_{\text{author}} = 34{,}596{,}581$; see the data flow in Fig. 1). In the dataset we used, if there were several country affiliations in the author's publications in the first year, the country with the largest number of publications was selected as the country of origin; authors with equal publication numbers in two or more countries were removed from the gender disambiguation analysis. Only authors with the first and last names available were passed through the NamSor API to retrieve a gender probability score. Authors without their first names were removed from the analysis, and all last-name variants were used. A gender probability score (the natural log of the ratio of probabilities, as determined by a naïve Bayes model, of the name receiving the classification of either male or female) greater than or equal to 0.85 ($N_{\text{author}} = 21{,}508{,}029$) was selected to ensure high-quality gender classification. This threshold has been described by Elsevier as a value that returns high evaluation metrics (Elsevier, 2020). Name-country combinations that fell short of this threshold were removed from the analysis. We used the Genderize.io gender determination tool in our previous national study of gender homophily, in which Polish scientists (with administratively provided gender) had international collaborators for whom gender had to be established (Kwiek & Roszka, 2021b). We used Genderize.io to refer to lists of first names and countries of origin to calculate the probability that the author's first name is feminine or masculine in the country of origin.

Second, to classify the author into a discipline, the dominant discipline from the disciplines assigned to the journals of all cited papers (i.e., all references) in the author's lifetime publishing portfolio was used. For this purpose, a table of publications was selected. All papers cited by the author ($N_{\text{cited reference}}=2{,}092{,}766{,}869$) were collected from the author's publication portfolio of any type (see dataflow in Fig. 1). Each cited paper had assigned disciplines that came from various sources (e.g., journals). Disciplines assigned to sources were based on four-digit All Science Journal Classifications (ASJC) codes. To switch to two-digit codes, the first two digits of the 4-digit value were selected. To avoid repetition of the same two-digit discipline at the level of the cited paper, only unique values were selected. The next step was to count for each author how many times they cited a paper from a given discipline. Assigning the dominant discipline to an author involved selecting the discipline for which the number of cited references was the highest (modal value). Authors who failed to be assigned a discipline or who had two or more dominant disciplines were removed from the sample ($N_{\text{included}}=32{,}794{,}309$; $N_{\text{removed}}=12{,}862{,}447$). We did not apply the random selection of a discipline in these cases. The set of authors and their disciplines was then narrowed down to only those who had a discipline from the STEMM field ($N_{\text{included}}=29{,}927{,}584$; $N_{\text{excluded}}=2{,}866{,}725$).

Third, to classify an author into a country, the dominant country indicated in the author's publication portfolio was used. For this purpose, a publication table was selected. From the publication portfolio of an author (publications of any type), all the affiliation countries indicated by the author were collected. Then, for each author, the number of times they indicated a country was counted. The assignment of the dominant country was based on selecting the country for which the number of publications of any type was the highest (modal value). As in the case of determining disciplines, authors who failed to be assigned to a country or had two or more dominant countries were removed from the sample ($N_{\text{inclued}}=39{,}405{,}552$; $N_{\text{excluded}}=6{,}251{,}204$). The set of authors and their countries was then narrowed down to 38 OECD countries only ($N_{\text{inclued}}=23{,}619{,}928$, $N_{\text{excluded}}=15{,}785{,}624$).

An important point when discussing the assignment of scientists to their modal countries is international mobility. Our study covers a period of up to 23 years of individual careers, and all we know about scientists changing countries or remaining in a single country comes from their publication data (see ESM, "International mobility rates"). Although mobility can be discussed in much detail in small-scale studies using different data sources, especially CVs, in large-scale studies in practical terms, our knowledge of international mobility can come only from bibliometric datasets (see, e.g., Sanliturk et al., 2023, who examined the mobility of scholars based on 36 million publications in Scopus).

Finally, to assign authors to the year of exit from publication, the tables of publications were selected. For each author, their identifiers from the Scopus database and the year of the last publication of any type are indicated. Then, the year following the year of the last publication was selected as the year of exit from publishing (the year of the last publication + 1). To classify an author as a censored observation, 2019 was selected as the last year of the study. Any author for whom the year of exiting publication was after 2019 (2020, 2021, 2022) was classified as a censored observation.

To estimate the magnitude of the error in the case of a potential return to publication by scientists after 2022 (for which years we do not yet have the data), we conducted analyses of the median frequency of publishing by scientists on an annual basis for the two cohorts studied. For the 2000 cohort, the median publication frequency for 86.70% of men and 83.66% of women (see Supplementary Table 3) was at least one publication annually, and for only 1.10% of men and 1.47% of women, the publication frequency was every fourth year. A zero-year break means publishing at least one article every year, a 1-year break means publishing every second year, a 2-year break means publishing at least one article every third year, and so forth. Additionally, for the 2010 cohort (Supplementary Table 4), for only 2.85% of men and 3.36% of women, the publication frequency is every fourth year. In both cases, there are large disciplinary variations. Men and women may be publishing with smaller intensity (generally, women tend to be less productive than men in STEMM disciplines, partially because of smaller collaboration networks)—but their publishing frequency per year is similar. In our attrition-focused study, frequency is more consequential than productivity.

Our study has an OECD focus: We analyze the science profession through cohorts of scientists in 38 OECD countries combined across time and STEMM disciplines. However, it is also possible to compare the differences by country in an interactive format where we provide the results of the Kaplan–Meier probability by country, discipline, and gender for all 11 cohorts (the 2000–2010 cohorts; see "Special case, the USA" in ESM). The snapshot in Fig. 2 shows the probability of staying in science for women from the 2000 cohort of scientists after 10 years for major European OECD member states in MED: It reaches 42% in Germany, as opposed to as much as 69% in Poland and 72% in Portugal.

## Limitations

The present study is not without limitations. Our research is clearly focused on publishing scientists only (the nonpublishers, or scientists with zero-level publishing, do not appear in our analysis); we examine publishing scientists of any sector rather than higher education personnel of all ranks, which is found in traditional accounts of gendered attrition in science (Deutsch & Yao, 2014; Goulden et al., 2011; Kaminski & Geisler, 2012; White-Lewis et al., 2023; Zhou & Volkwein, 2004).

Importantly, the conceptualization of leaving science as stopping publishing does not entail any other academic roles, such as teaching or administration, or any nonacademic roles, such as work in individual firms, corporations, or governments, even if prior research experience is deemed vital for these career paths. Here, "exit from science is a slippery concept since 'in science' and 'out of science' are not easily defined terms," with porous boundaries separating science from nonscience jobs (Preston, 2004, p. 14).

We discuss the entirety of academic life cycles (Kwiek & Roszka, 2024), from entering to leaving science, through a proxy for publishing the first and last scholarly publications indexed in the Scopus database. Hence, a sequence of publications replaces a sequence of much wider cognitive and social processes encompassing the various dimensions of doing science (Sugimoto & Larivière, 2023), with the assumption that not publishing in scholarly journals means not doing science anymore. Therefore, our representations of scientific careers are necessarily simplified: We actually examine publishing careers (Kwiek & Roszka, 2023). Our representation of scholarly output is necessarily reduced to globally indexed publications. In individual-level lifetime publication portfolios for every scientist in our dataset, nonindexed publications (and most publications in languages other than English) are not counted. As a result, in our research, the breadth of scientists' activities in academia (such as mentoring students, refereeing papers, reviewing grant proposals, and editing journals) is ignored (see e.g., Liu et al., 2023).

In other words, in the present research, active participation in science is defined through publishing; consequently, not publishing is defined as leaving science, in accordance

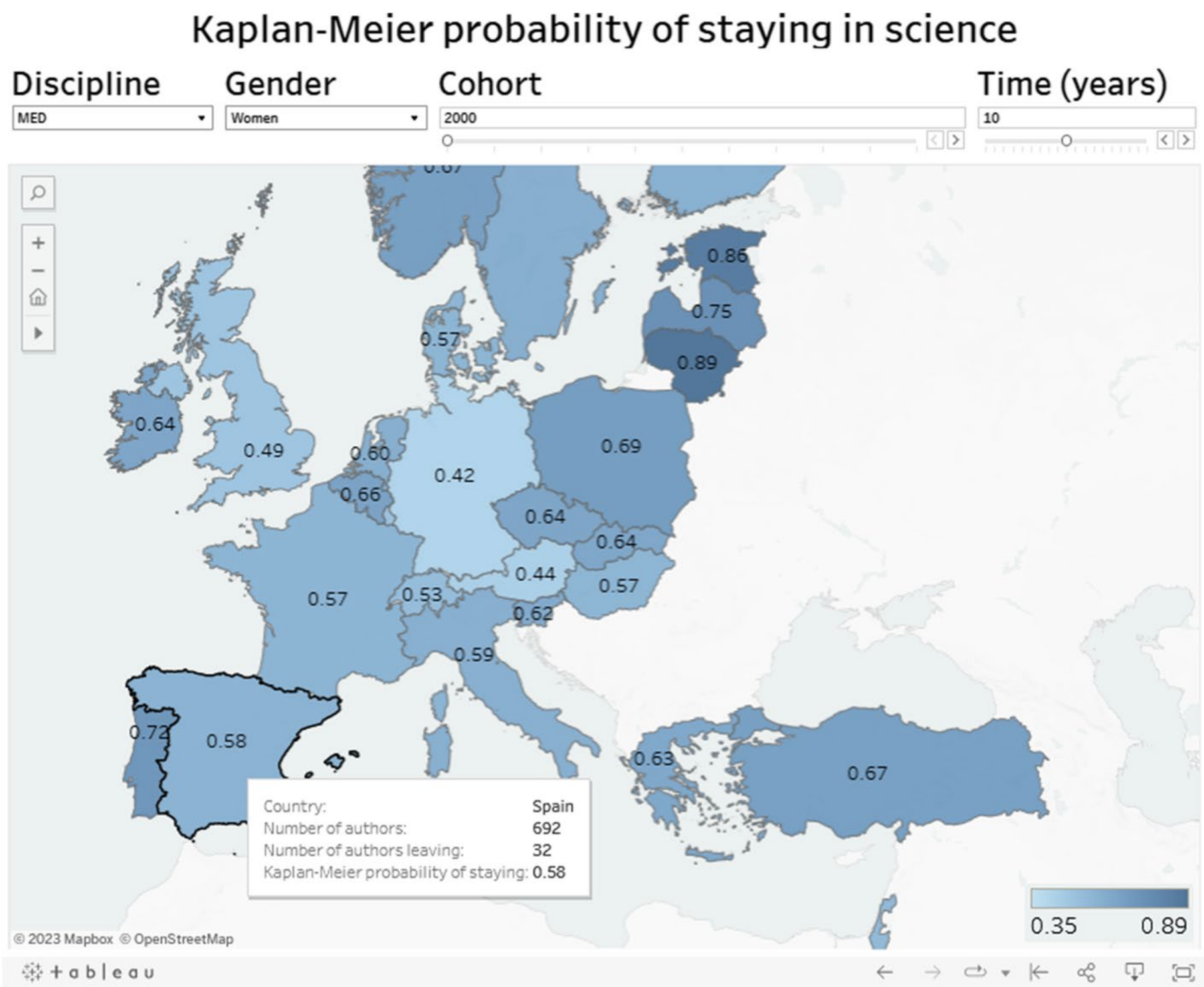

**Fig. 2** A snapshot of an interactive map (available from https://public.tableau.com/app/profile/marek.kwiek/viz/Attrition-in-science-OECD/Dashboard) in which full Kaplan−Meier probabilities of remaining in science (i.e., continuing publishing) are provided by country, discipline, gender, and for 11 cohorts from 2000 to 2010 ($N$ = 2,127,803 scientists, of which 1,289,756 are identified as men and 838,047 as women)

with the Mertonian tradition of the sociology of science (Kwiek, 2019). Nonpublishers in STEMM can continue their work in the academic sector but in other academic roles; however, currently, it is not possible to verify their intrasectoral or extrasectoral employment at the global level using bibliometric datasets (which might be possible using Large Language Models).

## Results

First, we tracked 142,776 STEMM scientists from the 2000 cohort until they stopped publishing (or until 2022). Figure 3 presents the Kaplan–Meier survival curve for all the STEMM disciplines combined, with the proportion of scientists who have lived in science (and the probability of staying in science, that is, of continuing to publish) shown on the y-axis and the number of years spent publishing since 2000 on the X-axis (with small crosses for right-censored cases for the three most recent years of 2020–2022).

The largest declines in the curve for both men and women appear at years 1, 2, 3, and 4 (which are generally years spent in doctoral schools). From year 4 onward, the difference between men and women becomes apparent: Every year starting with year 4, the proportion of surviving men is greater than the proportion of surviving women—which generally follows the patterns known from past research. Simplifying the results, approximately one-third of the 2000 cohort of scientists left science after 5 years, approximately half after 10 years, and approximately two-thirds by the end of the period examined (after 19 years), with the share of the leavers being consistently lower for men and higher for women. Thus, women are approximately one-tenth more likely to drop out of science than men after both 5 and 10 years (12.54% and 11.52%, respectively), and women are 6.33% more likely to drop out at the end of the studied period. After 19 years (only uncensored observations), the Kaplan–Meier probability of staying for women is 0.294 (29.4% of women from the original cohort continue publishing); for men, in contrast, it is considerably higher, reaching 0.336 (33.6%).

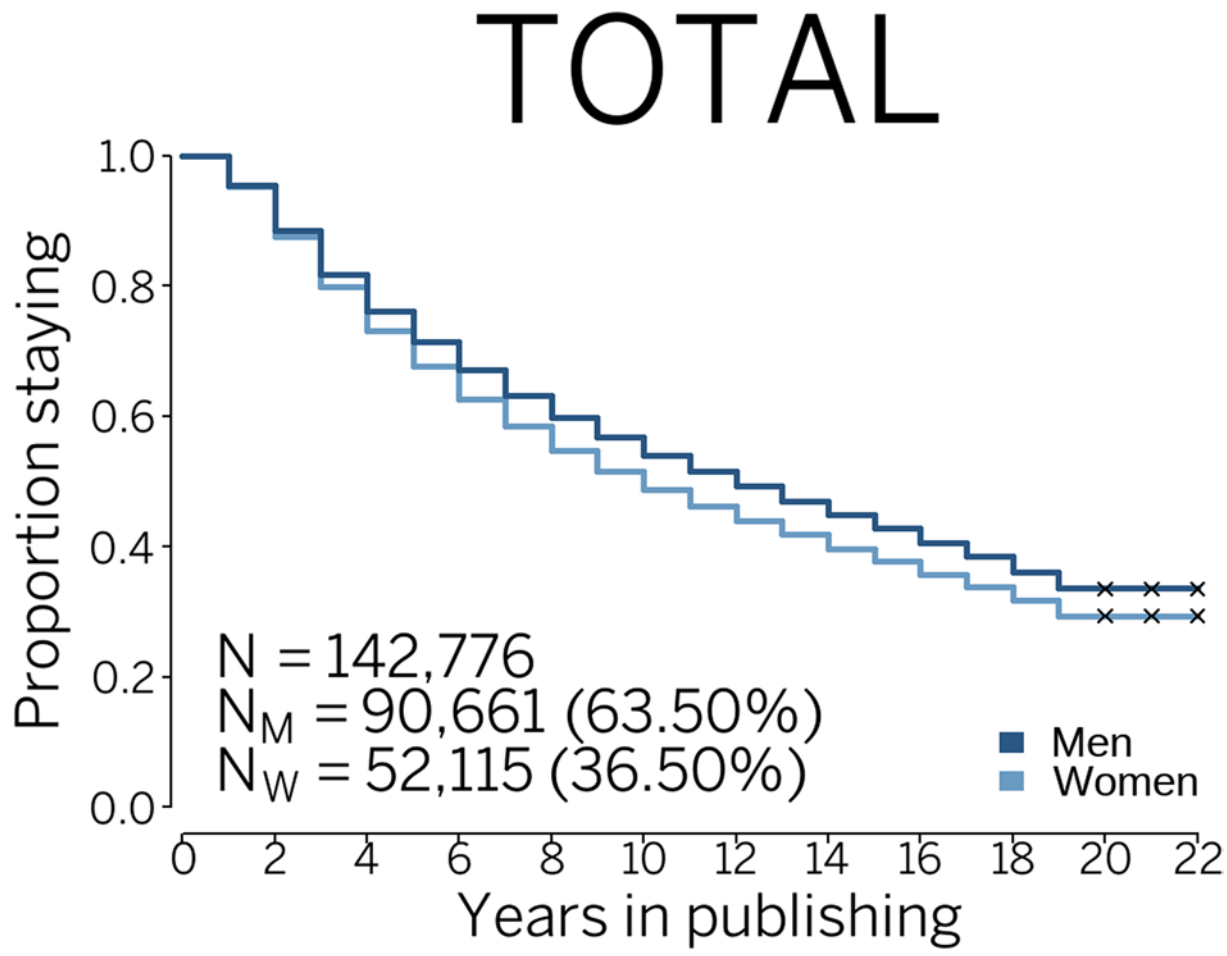

**Fig. 3** Kaplan–Meier survival curves according to gender (all disciplines combined) for the 2000 cohort of scientists ($N = 142{,}776$)

However, this aggregated general picture of attrition in science for all STEMM disciplines combined obscures the different disaggregated pictures for particular disciplines, with substantial cross-disciplinary variation (Fig. 4).

For the 2010 cohort, the Kaplan–Meier survival curves for men and women were drastically different (compare the 2000 cohort in Fig. 4 and the 2010 cohort in Supplementary Fig. 1); for all disciplines combined, they were nearly identical for men and women. They are also nearly identical for math-intensive COMP, PHYS, and ENG.

The details of the Kaplan–Meier estimate for the 2000 cohort population (Table 2) show that, for women, the probability of staying in science in the first year is 95.1% (95% confidence interval: 95.0–95.3%), in the second year is 87.5%, and in the fourth year is 73.1%; thus, the cumulative probability of leaving after 4 years is 26.9%. For women, the probability of staying in science after 5 years is 67.7%, after 10 years is 48.7%, and at the end of the study period (i.e., after 19 years) is 29.4%. In contrast, for men, the corresponding percentages are 71.3%, 54.0%, and 33.6%, respectively, which are significantly greater for each period studied.

The year-by-year data indicate that both the numbers and percentages of men and women leaving science are greatest in the first 6 to 8 years, and the Kaplan–Meier probability of staying from year to year is generally greater in the second decade of publishing than in the first decade of publishing. The Kaplan–Meier probability of staying is lower than 50% for women in year 10 and for men in year 12. The estimated probability that a woman will survive in science for 15 years or more is 37.7%, as opposed to 42.9% for men (see Supplementary Tables 6–8 for median exit time from science).

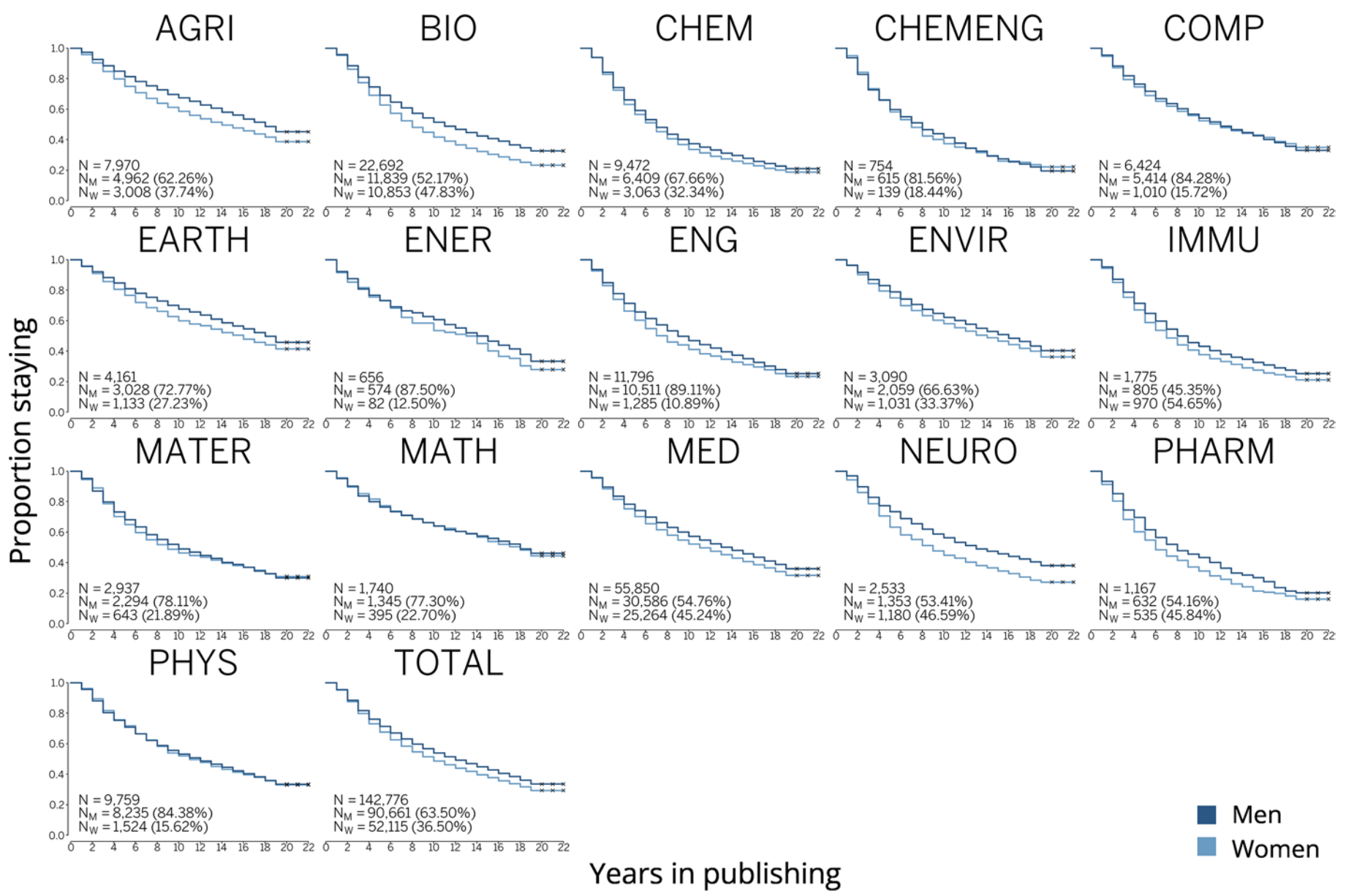

**Fig. 4** Kaplan–Meier curves by discipline and gender for the 2000 cohort of scientists ($N = 142{,}776$)

For the 2010 cohort (Table 3), 9 years after entering science, the probability of remaining for women was 0.414 (41.4% of women from the original cohort continue publishing); for men, the probability of staying was only slightly greater at 0.424 (42.4%), which is a dramatic lack of difference compared with the results for the 2000 cohort, where the results were substantially gender sensitive. In eight disciplines, including the math-intensive COMP, ENG, MATH, and PHYS, the statistical tests showed that there were no statistically significant differences in the survival curves for men and women; however, for all disciplines combined, as well as for the largest disciplines of MED, BIO, and AGRI, with women representing approximately 50%, the differences were statistically significant (see Supplementary Table 10, six methods).

Our special interest in scientific careers is in disciplines with the largest and smallest (or none at all) differences in survival curves between men and women. In the two largest disciplines and two disciplines with the largest number of women, medicine (MED) and biochemistry, genetics, and molecular biology (BIO), the curves for men and women are clearly divergent (and the difference is statistically significant).

Focusing on BIO (Fig. 5, left panel, contrasted with physics and astronomy, PHYS, right panel), with 47.83% of women and 22,692 scientists in the cohort examined, the largest declines are in years 2, 3, and 4. Starting in year 3, there is an ever-increasing difference in men–women between the two survival curves, here showing smoother declines.

For women in BIO, the probability of leaving science after 5 years is 37.2%, after 10 years is 58.3%, and at the end of the study period (i.e., after 19 years) is 76.6%. For men, the corresponding figures are strikingly lower: 30.8%, 48.6%, and 67.3%, respectively. Thus, women are as much as one-fifth more likely to drop out of science than men after both 5 and 10 years (20.78% and 19.96%, respectively), and 13.82% are more likely to drop out at the end of the studied period.

**Table 2** Kaplan–Meier estimate for the 2000 cohort of scientists by gender (all disciplines combined) with total counts for men and women, time (in years), number of observations of scientists leaving science, and Kaplan–Meier probability of staying with a 95% confidence interval

| Time (years) | Women (cohort 2000) | | | Men (cohort 2000) | | |
|---|---|---|---|---|---|---|
| | *N* | Leaving science (*N*) | KM probability (staying) with 95%CI and SE | *N* | Leaving science (*N*) | KM probability (staying) with 95%CI and SE |
| 1 | 52,115 | 2530 | 0.951 (0.950–0.953)[1] | 90,661 | 4151 | 0.954 (0.953–0.956)[1] |
| 2 | 49,585 | 3985 | 0.875 (0.872–0.878)[1] | 86,510 | 6302 | 0.885 (0.883–0.887)[1] |
| 3 | 45,600 | 3948 | 0.799 (0.796–0.803)[2] | 80,208 | 6114 | 0.817 (0.815–0.820)[1] |
| 4 | 41,652 | 3553 | 0.731 (0.727–0.735)[2] | 74,094 | 5062 | 0.761 (0.759–0.764)[1] |
| 5 | 38,099 | 2838 | 0.677 (0.673–0.681)[2] | 69,032 | 4356 | 0.713 (0.710–0.716)[2] |
| 6 | 35,261 | 2602 | 0.627 (0.623–0.631)[2] | 64,676 | 3934 | 0.670 (0.667–0.673)[2] |
| 7 | 32,659 | 2183 | 0.585 (0.581–0.589)[2] | 60,742 | 3458 | 0.632 (0.629–0.635)[2] |
| 8 | 30,476 | 1961 | 0.547 (0.543–0.551)[2] | 57,284 | 3110 | 0.598 (0.594–0.601)[2] |
| 9 | 28,515 | 1665 | 0.515 (0.511–0.520)[2] | 54,174 | 2774 | 0.567 (0.564–0.570)[2] |
| 10 | 26,850 | 1472 | 0.487 (0.483–0.491)[2] | 51,400 | 2465 | 0.540 (0.537–0.543)[2] |
| 11 | 25,378 | 1264 | 0.463 (0.458–0.467)[2] | 48,935 | 2225 | 0.515 (0.512–0.518)[2] |
| 12 | 24,114 | 1158 | 0.440 (0.436–0.445)[2] | 46,710 | 2055 | 0.493 (0.489–0.496)[2] |
| 13 | 22,956 | 1151 | 0.418 (0.414–0.423)[2] | 44,655 | 2032 | 0.470 (0.467–0.473)[2] |
| 14 | 21,805 | 1089 | 0.398 (0.393–0.402)[2] | 42,623 | 1889 | 0.449 (0.446–0.453)[2] |
| 15 | 20,716 | 1048 | 0.377 (0.373–0.382)[2] | 40,734 | 1884 | 0.429 (0.425–0.432)[2] |
| 16 | 19,668 | 1033 | 0.358 (0.353–0.362)[2] | 38,850 | 1959 | 0.407 (0.404–0.410)[2] |
| 17 | 18,635 | 1002 | 0.338 (0.334–0.342)[2] | 36,891 | 2020 | 0.385 (0.381–0.388)[2] |
| 18 | 17,633 | 1064 | 0.318 (0.314–0.322)[2] | 34,871 | 2070 | 0.362 (0.359–0.365)[2] |
| 19 | 16,569 | 1228 | 0.294 (0.290–0.298)[2] | 32,801 | 2350 | 0.336 (0.333–0.339)[2] |

[1]Standard error of 0.001

[2]Standard error of 0.002

PHYS, in contrast, is a perfect example of the lack of gender differences in attrition. For women in PHYS, the probability of leaving science after 5 years is 28.1% (and 29.2% for men), after 10 years is 47.9% (46.9% for men), and at the end of the study period (i.e., after 19 years) is 66.9% (66.5% for men).

Strikingly, in the three math-intensive disciplines of MATH, COMP, and PHYS (the fourth, ENG, being an exception), which have very low numbers and percentages of women, the Kaplan–Meier survival curves for men and women are nearly identical (see PHYS in Fig. 5, right panel), with the two curves almost overlapping. As confirmed for 38 OECD countries by a large sample of physicists and astronomers who all started publishing in 2000, gender differences in attrition in PHYS do not exist. Overall, for the four math-intensive disciplines, the differences are not statistically significant; for the disciplines with the highest proportions of women (AGRI, BIO, and MED, as well as all disciplines combined), the differences are statistically significant (see Supplementary Table 5 for statistical tests for the difference between men and women, six methods).

To obtain a more comprehensive view of attrition and retention in science, we correlated the data on men and women from the 2000 cohort (Fig. 6) and the 2010 cohort (Fig. 7) to produce survival regression curves (Panel A), hazard rate curves (Panel B), and kernel density curves (Panel C) to tell the story of attrition and retention from different perspectives.

**Table 3** Kaplan−Meier estimates for the 2010 cohort of scientists by gender (all disciplines combined) with total counts for men and women, time (in years), number of observations of scientists leaving science, and Kaplan−Meier probability of staying with a 95% confidence interval

| | Women (cohort 2010) | | | Men (cohort 2010) | | |
|---|---|---|---|---|---|---|
| Time (years) | *N* | Leaving science (*N*) | KM probability (staying) with 95%CI and SE | *N* | Leaving science (*N*) | KM probability (staying) with 95%CI and SE |
| 1 | 97,145 | 5030 | 0.948 (0.947–0.950)[1] | 135,698 | 7375 | 0.946 (0.944–0.947)[1] |
| 2 | 92,115 | 8686 | 0.859 (0.857–0.861)[1] | 128,323 | 12,183 | 0.856 (0.854–0.858)[1] |
| 3 | 83,429 | 8090 | 0.776 (0.773–0.778)[1] | 116,140 | 11,164 | 0.774 (0.771–0.776)[1] |
| 4 | 75,339 | 7369 | 0.700 (0.697–0.703)[1] | 104,976 | 9869 | 0.701 (0.698–0.703)[1] |
| 5 | 67,970 | 6470 | 0.633 (0.630–0.636)[2] | 95,107 | 8552 | 0.638 (0.635–0.640)[1] |
| 6 | 61,500 | 5904 | 0.572 (0.569–0.575)[2] | 86,555 | 7851 | 0.580 (0.577–0.583)[1] |
| 7 | 55,596 | 5499 | 0.516 (0.513–0.519)[2] | 78,704 | 7322 | 0.526 (0.523–0.529)[1] |
| 8 | 50,097 | 4984 | 0.464 (0.461–0.468)[2] | 71,382 | 7078 | 0.474 (0.471–0.477)[1] |
| 9 | 45,113 | 4929 | 0.414 (0.411–0.417)[2] | 64,304 | 6745 | 0.424 (0.422–0.427)[1] |

[1]Standard error of 0.001

[2]Standard error of 0.002

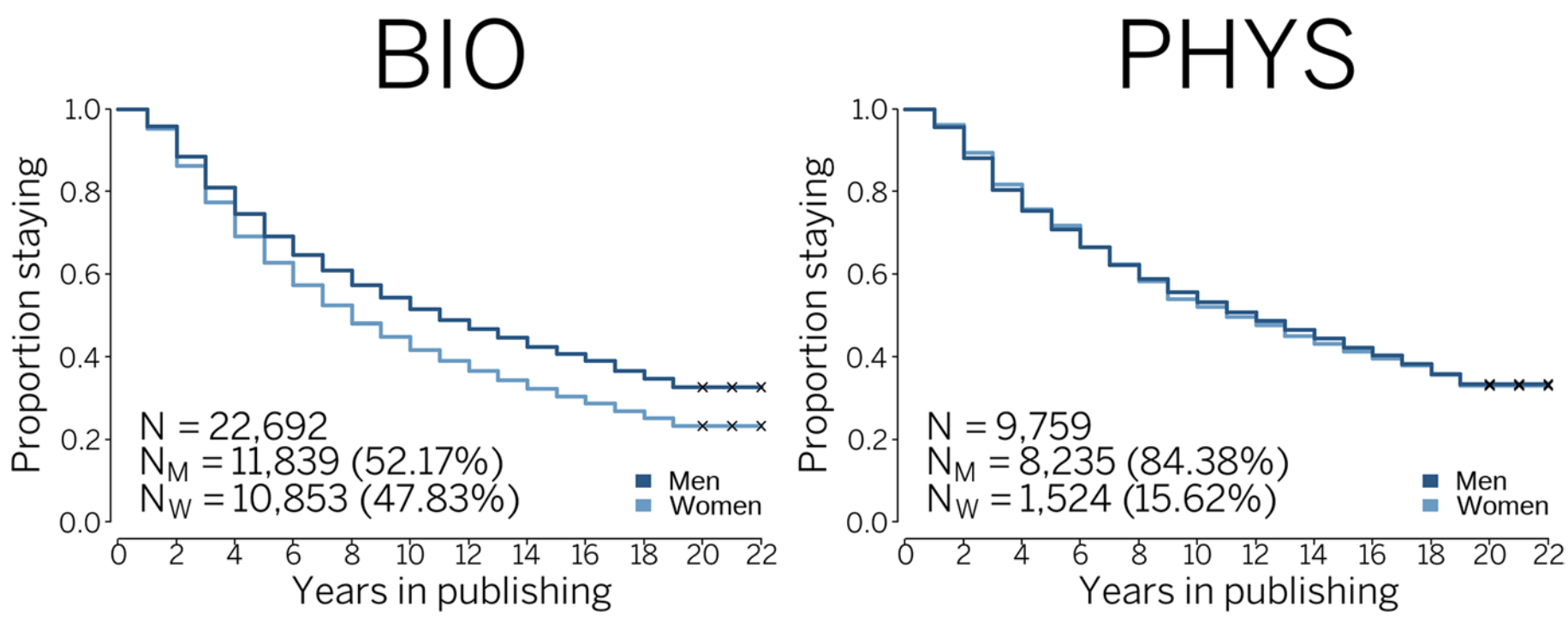

**Fig. 5** Kaplan–Meier survival curves according to gender; BIO (*N*=22,692; left panel) vs. PHYS (*N*=9759; right panel) in the 2000 cohort of scientists

In general, survival analysis models the time until an event of interest (here, ceasing publishing) occurs. For all scientists from the 2000 and 2010 cohorts, there is the same starting time (year 2000, year 2010), and for some proportion of the cohort, an event occurs after some time (in years passed since 2000 and 2010, respectively). The hazard is the probability of an event occurring during any given time point within the period studied. We calculate the probability that any given scientist will cease publishing each year and plot the results.

The survival regression function models the Kaplan−Meier estimation, which shows the changing proportion of scientists who are still publishing in subsequent years. The survival

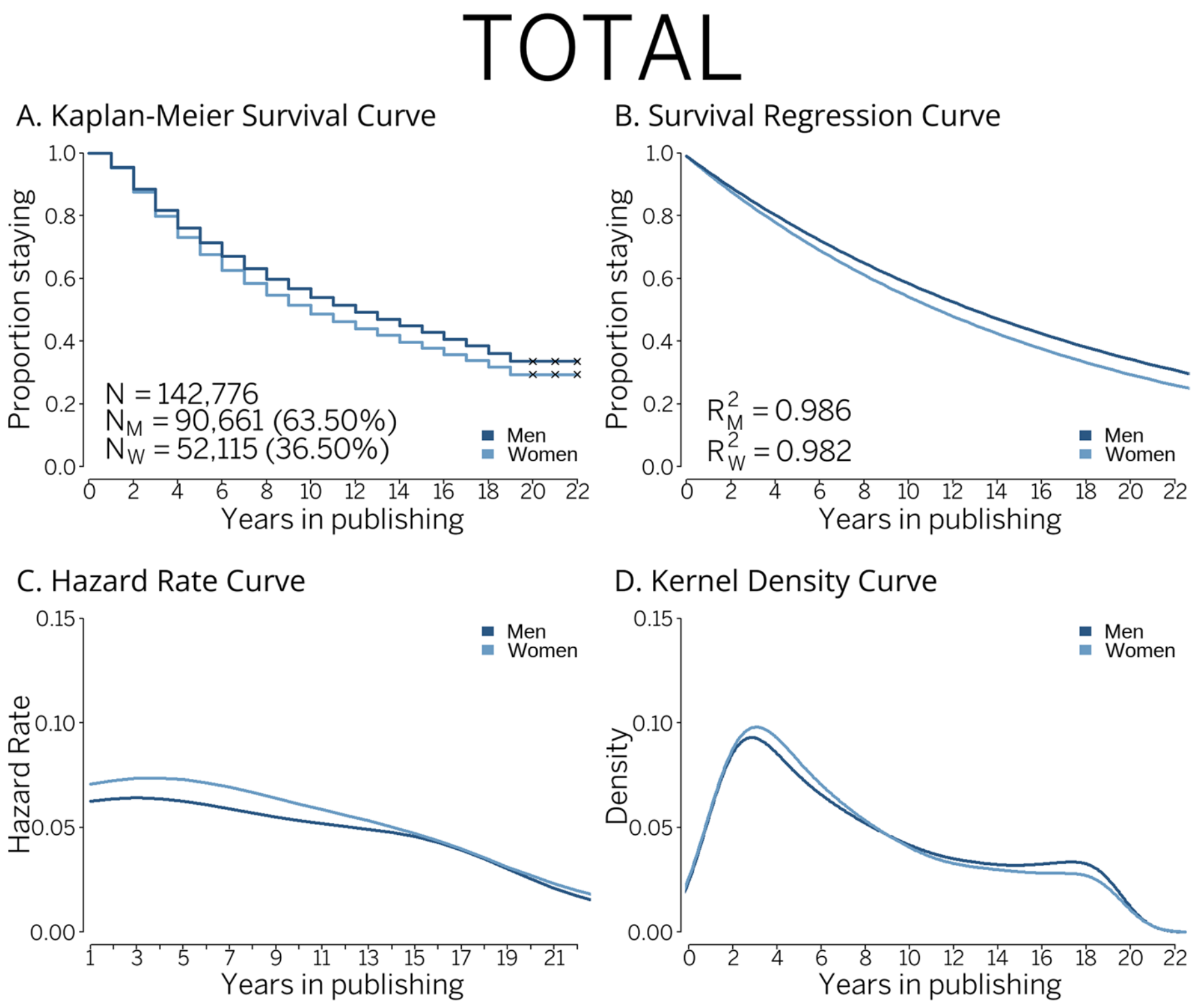

**Fig. 6** Kaplan–Meier survival curve, survival regression curve (exponential distribution fitting of Kaplan–Meier curve), hazard rate curve (B-splines smoothing method, smoothing parameter 10 k), and kernel density curve (B-splines smoothing method, bandwidth 2, component per point based on Gaussian curve), all disciplines combined, the 2000 cohort of scientists ($N$ = 142,776)

regression curve is actually a smoothed Kaplan–Meier curve. It starts with 1 on a y-axis because all scientists are publishing at time zero before then gradually declining to some level for men and women.

Consequently, similar to the Kaplan–Meier survival curves, the survival regression curves for all disciplines combined for both cohorts (Figs. 6b and 7b) indicate a much steeper decline for men and women in the early years of their publishing careers and a much smoother decline in later years of their publishing careers, with the increasing divergence between the curves for men and women from the 2000 cohort. For the 2000 cohort, as opposed to the 2010 cohort, the surviving men stay in science at a higher rate than the surviving women, and both men and women later in their careers stay in science at a higher rate than earlier in their careers.

The hazard rate curves (panel C) provide another view of attrition: The rate of attrition for both men and women (for all disciplines combined) is greater in the early publishing years, and it is considerably lower in later years; for the 2000 cohort and for the first 15 years, women have greater chances of leaving science than men. The peak for men is in year 3, and for women, it is in year 4. Hazard rate curves show the probability of observing the event of interest in a given unit of time (here, year). The shape of the hazard rate curve provides important information about attrition and retention in science. The hazard

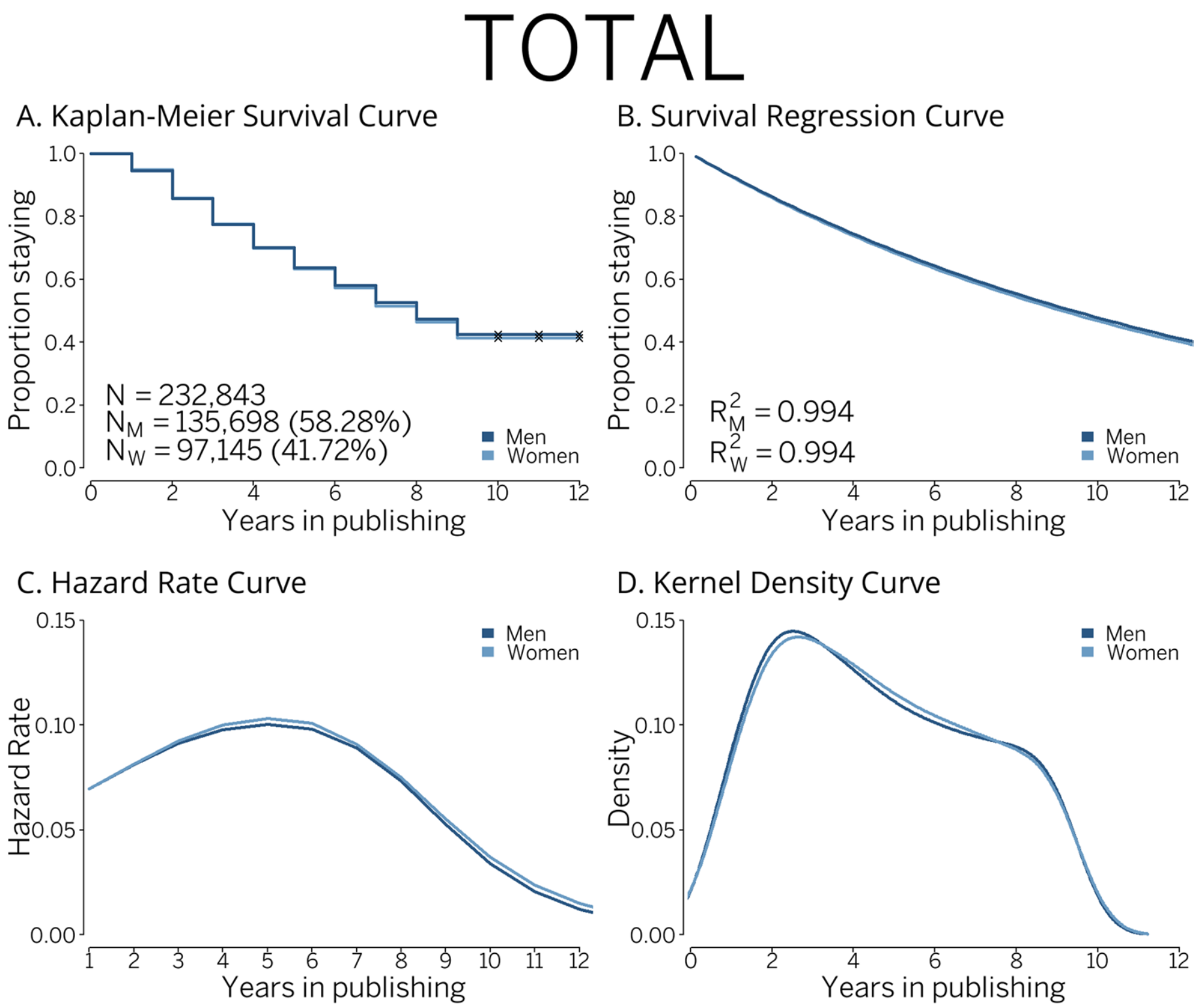

**Fig. 7** Kaplan–Meier survival curve, survival regression curve (exponential distribution fitting of Kaplan–Meier curve), hazard rate curve (B-splines smoothing method, smoothing parameter 10 k), and kernel density curve (B-splines smoothing method, bandwidth 2, component per point based on Gaussian curve). All disciplines combined, the 2010 cohort of scientists ($N$ = 232,843)

functions shown for the 2000 cohort are monotonically decreasing: The short-term risk of ceasing publishing is greater in the early years of publishing careers, after which the risk decreases over time. The riskiest years are the early years in science. The short-term risk never increases again to the levels noted for the early years. Later years in science are less risky than early years, although the conditions under which scientists cease publishing differ in different career stages. However, for the 2010 cohort, the risk increased until year 6 and then decreased over time. Most importantly, there are no gender differences, and the curves for men and women overlap.

Finally, kernel density plots use kernel density estimation to create a smoothed, continuous curve that approximates the underlying data distribution. Kernel density estimations have been used to estimate the probability density function of scientists' years since the first publication. Kernel density estimations display the distribution of values in a dataset using one continuous curve—they identify the shape of the distribution (left and right skewness; symmetry; unimodal or multimodal distributions based on the number of peaks, etc.). They work better than histograms in displaying the shape of a distribution because their shape is not affected by the number of bins used or by dramatic differences between them (no bin width decisions need to be made); they can be flexibly used to compare distributions of two or more datasets.

Kernel density curves (panel D) show how all men and women who actually left science from the 2000 cohort and from the 2010 cohort are distributed over time (or among the years since 2000 and since 2010). The density curves are easily interpretable because the area under the curve always reaches 100%. For the 2000 cohort, the shares of scientists who left science are the highest for the first 3–6 years, and for the 2010 cohort, the critical years are the first 2–5 years. Both men and women from both cohorts of scientists are more likely to leave science early on after starting their publishing careers than later. Once they have survived the first 10 years, their likelihood of leaving is much lower than in their early years.

This generalized story for all disciplines combined obscures a number of much more nuanced discipline-specific stories. The Kaplan–Meier curves for BIO and PHYS analyzed for the 2000 cohort tell fundamentally different stories. These findings are confirmed when the survival regression curves are compared (BIO and PHYS, Fig. 8A), with similar higher probabilities of leaving in early years and lower probabilities in later years—but with stark differences between men and women in the two disciplines. Our data show that women in BIO disappear from science with the passage of time in ever-larger proportions compared with men; in contrast, women in PHYS disappear from science in almost exactly the same proportions as men in the entire period examined.

Additionally, hazard rate curves (Fig. 8B) tell a similar story in which, in BIO, the attrition rate for women is substantially higher than the attrition rate for men across all the years examined; in PHYS, in contrast, the hazard curve rates for men and women are almost identical. The shape of the hazard function may differ between groups of interest (as in BIO) or be identical (as in PHYS). The hazard for men is markedly lower than the hazard for women at all data points in BIO; in PHYS, in contrast, it is almost identical for all data points.

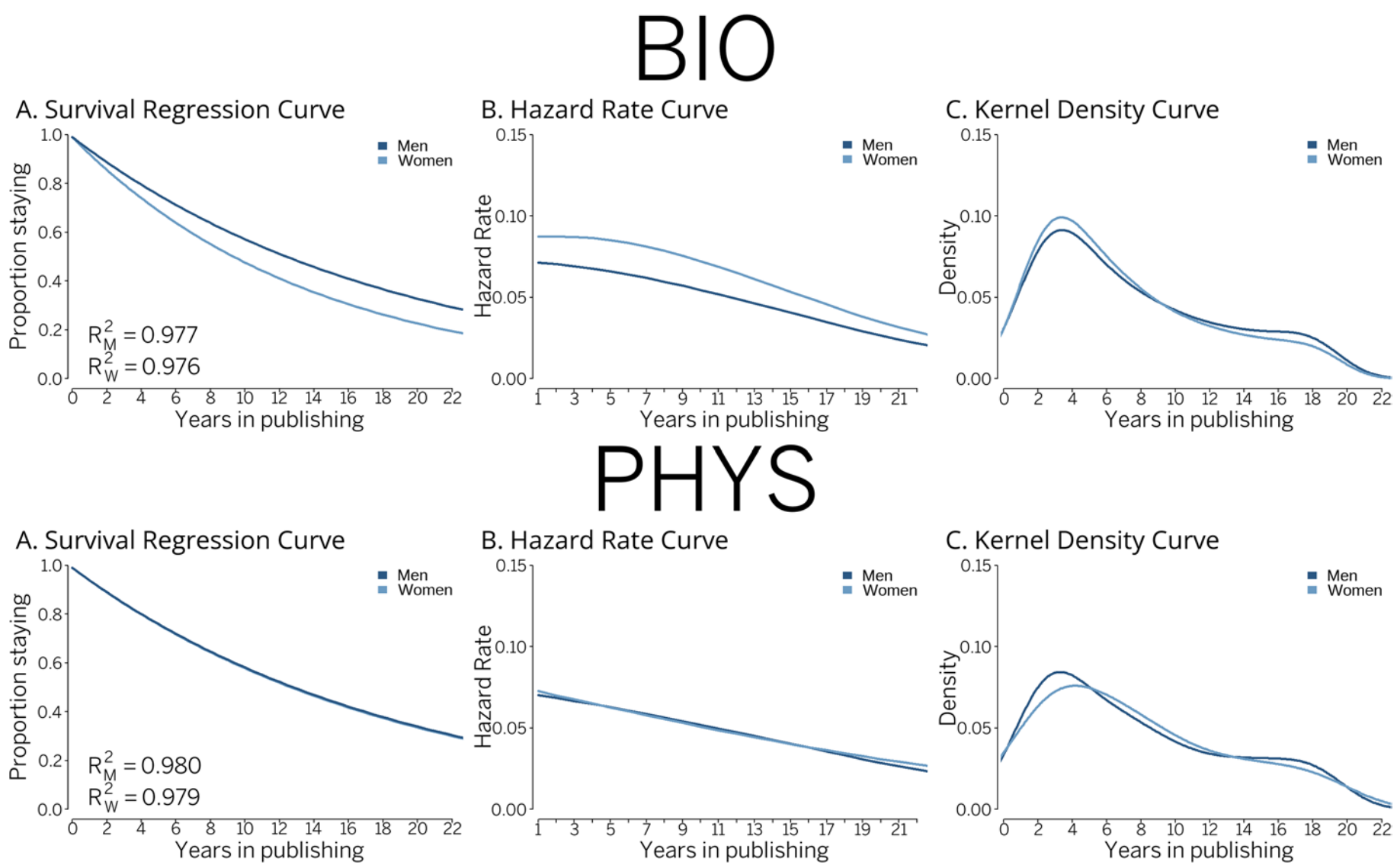

**Fig. 8** Survival regression curve (exponential distribution fitting of Kaplan−Meier curve), hazard rate curve (B-splines smoothing method, smoothing parameter 10 k), and kernel density curve (B-splines smoothing method, bandwidth 2, component per point based on Gaussian curve), BIO ($N$ = 22,692), and PHYS ($N$ = 9750), the 2000 cohort of scientists

Finally, kernel density curves for BIO and PHYS (Fig. 8C) clearly show similar intradisciplinary patterns for men and women (higher attrition in early years) and different cross-disciplinary patterns for men and women. In fact, although in BIO attrition for women is greater than for men in the early years, in PHYS in years 2 through 6, attrition is greater for men. This is a surprising finding, which is also confirmed for COMP and MATH (see Supplementary Figs. 2 and 3).

From a cross-cohort comparative perspective, our analyses indicate a fundamental transformation in attrition patterns across disciplines (and in all disciplines combined). The differences between men and women, which are so starkly visible for the 2000 cohort, almost disappear for the 2010 cohort.

This finding has important potential research and policy implications in view of the vast literature on attrition and retention in academic science. The science sector globally seems to be undergoing powerful transformations, and the findings valid for older cohorts of scientists (here, the 2000 cohort) may not be applicable to younger cohorts (here, the 2010 cohort). Time in science—in this case, the difference of about a decade—matters.

The Kaplan–Meier survival curves for all disciplines combined take vastly different forms for the two cohorts: For the 2010 cohort (Fig. 7), first, attrition rates are much higher and decline much more dramatically (50% or more of both men and women disappear in year 8), and there are no gender differences at all. This finding is confirmed by the shapes of the survival regression curves for the two cohorts, again with no gender difference; the confirmation comes from the hazard rate curves, which are nearly identical for men and women, and from the kernel density curve, which testifies to the generally similar distribution of scientists who left science over time. Specifically, the kernel density distribution shows dramatically high attrition rates for both men and women in the first 4 years of staying in science (Fig. 7B, C, and D).

Again, for the 2010 cohort, a large picture of all combined disciplines hides behind smaller and complicated discipline-specific pictures. In some disciplines, the changes are much more fundamental than in others. We restrict our cursory view to the two disciplines selected for cohort 2000 (BIO and PHYS) and for COMP, which is a special case. In BIO, the differences in the Kaplan–Meier survival curves and in the survival regression curves for men and women are considerably smaller for the 2010 cohort than for the 2000 cohort; both hazard rate curves and kernel density curves for men and women show similar differences, albeit with different intensities over the first 10 years. PHYS, in contrast, remains a discipline with minor (or even none at all) gender differences in attrition, with women showing a slightly greater probability of leaving science in early and late years than men. The kernel density distribution shows that men are much more likely to leave science than women in years 2–4; that is, men are more likely to leave science very early in their careers.

First, the survival regression curves for the two cohorts mirror the Kaplan–Meier curves for the two cohorts but are smoothed. For those disciplines for which there were differences between men and women for the 2000 cohort, these differences continue for the 2010 cohort but are smaller (see, e.g., the large disciplines of MED, BIO, and AGRI). For those disciplines where the differences were invisible—such as COMP and PHYS—they continue to be invisible. Overall, the attrition gaps between men and women have been closing across the board so that, for all disciplines combined, they are invisible.

Second, the hazard function curves by discipline for the two cohorts tended to confirm the findings based on Kaplan–Meier curves and survival regression curves. The chances of leaving science are much less gendered for the 2010 cohort than for the 2000 cohort for most disciplines, especially in the early years of publishing careers. The differences between men and women are especially stark in the 2010 cohort for BIO, NEURO, and PHARM (as in the case of the 2000 cohort). However, for the three math-intensive disciplines of COMP, MATH, and PHYS, the generally overlapping curves for the 2000 cohort are more volatile for the 2010 cohort over time and are still very similar for men and women.

The kernel density curves for the two cohorts largely tell the same story; specifically, should the 2010 cohort be observed for another decade (i.e., in 2032), the distributions of those who leave science within the two cohorts may be much more similar. The early years (years 2–6) are those when large proportions of scientists leave, but gender differences in attrition for the 2010 cohort are considerably lower than those for the 2000 cohort. The distributions for all disciplines combined are almost identical for men and women, and in several disciplines, more men leave early than women do (e.g., in CHEMENG, COMP, MATER, and PHYS). Overall, the patterns of distribution of the leavers for the 2010 cohort are much less gendered than those for the 2000 cohort, confirming the observation that, over time, the differences in attrition in science between men and women are ever less pronounced.

Finally, computing is a special case that has traditionally attracted attention in academic career studies (e.g., Branch & Alegria, 2016; Fox & Kline, 2016; Fox & Xiao, 2013; Fox et al., 2017) along with mathematics (Mihaljević & Santamaría, 2020; Mihaljević-Brandt et al., 2016; Watt et al., 2017). Participation in computing is especially challenging: In 38 OECD countries in 2021, 18.20% of women published in COMP, with the percentage only slightly higher for the youngest generation of publishing scientists (20.65%, publishing experience of no more than 5 years) and with 113 years needed for gender parity (50/50) and 78 years needed for gender balance (40/60), based on current participation trends (Kwiek & Szymula, 2023).

Our analyses for COMP sent a clear message (Figs. 9 and 10): For both cohorts, with 15.72% of women starting publishing in 2000 and 17.96% of women starting publishing in 2010, there were no gender differences in the Kaplan−Meier survival curves (and in their smoothed versions: survival regression curves). The hazard rate curves indicate that for the majority of years examined, the probability of leaving science is actually greater for men than for women, and kernel density plots show that the peak in the distribution of leavers is approximately year 4 of publishing careers.

## Discussion and conclusions

Attrition and retention in science are long-term processes and require large-scale multi-country and longitudinal datasets to study if we want to move beyond single countries and analyze the phenomena by discipline and over time. "Leaving science" is undergoing significant transformations as new cohorts of scientists enter science every year under new (working, professional, and other) conditions (see Milojevic et al., 2018).

Our cohort-based and longitudinal study revealed that, behind aggregated changes at the level of all the STEMM disciplines combined, widely nuanced changes occurred, with various intensities at the discipline level and over time. In our conceptualization of "ceasing publishing," "attrition in science" means different things for men and different things for women in different disciplines, and it means different things for scientists from different cohorts entering the scientific workforce.

Approximately one-third of the 2000 cohort of scientists left science after 5 years, approximately half after 10 years, and approximately two-thirds by the end of the period examined (after 19 years), with the share of the leavers being consistently lower for men

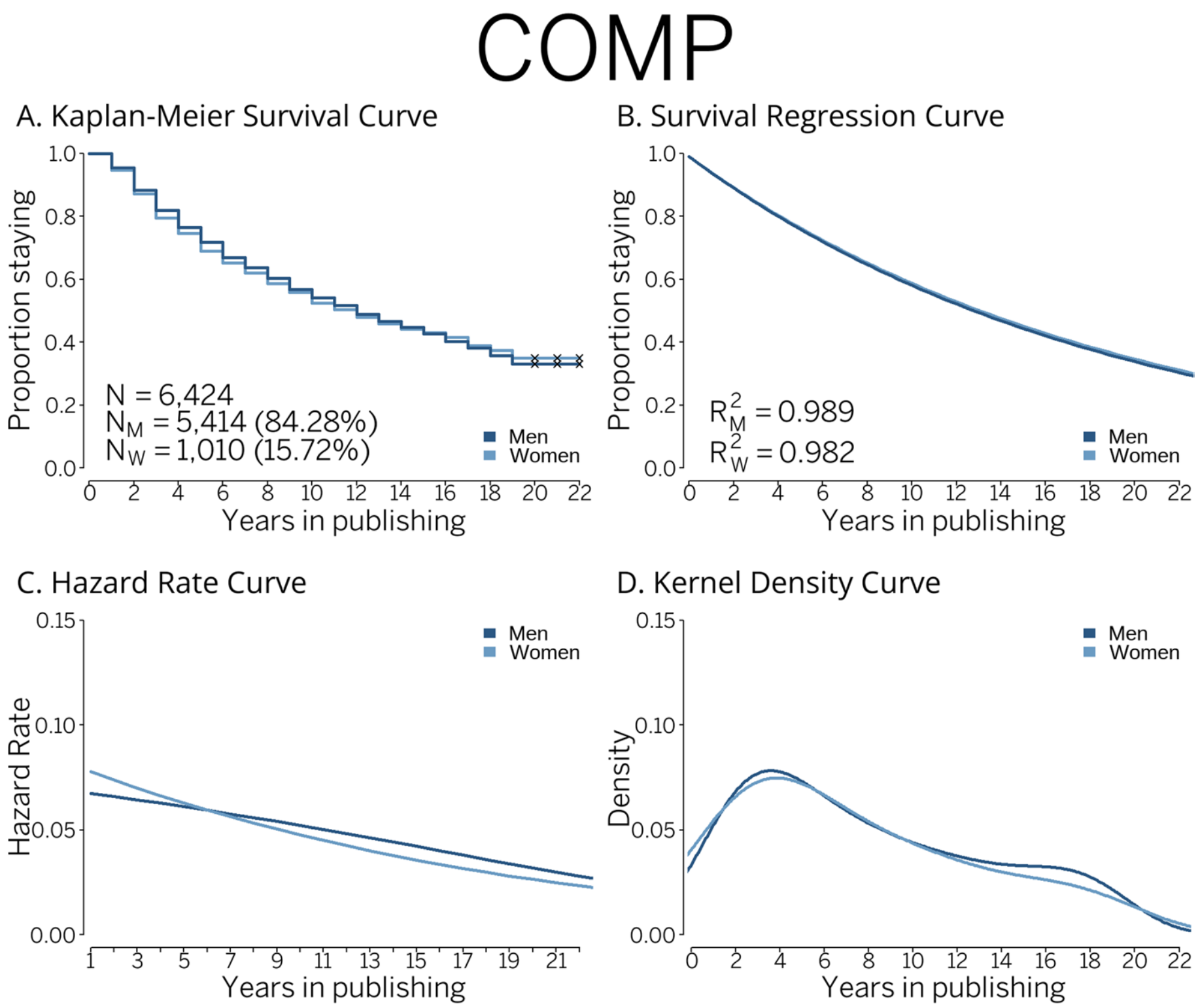

**Fig. 9** Kaplan−Meier survival curve, survival regression curve (exponential distribution fitting of Kaplan−Meier curve), hazard rate curve (B-splines smoothing method, smoothing parameter 10 k), kernel density curve (B-splines smoothing method, bandwidth 2, component per point based on Gaussian curve), COMP, and the 2000 cohort of scientists ($N = 6424$)

and higher for women. Women are approximately one-tenth more likely to drop out of science than men after both 5 and 10 years (12.54% and 11.52%, respectively), and women are 6.33% more likely to drop out at the end of the studied period.

However, gender differences in attrition are smaller for scientists entering science in 2010 than for scientists entering science in 2000: With more women in science and more women within cohorts, attrition is becoming less gendered. At the level of 38 OECD countries, gender differences clearly visible for the 2000 cohort for all disciplines combined disappeared for the 2010 cohort. The aggregated pictures hide behind them nuanced disaggregated pictures for both cohorts, with different ongoing processes for different disciplines.

The Kaplan–Meier curves for the two contrasted disciplines of biochemistry, genetics, and molecular biology (BIO, large, 47.83% of women) and physics and astronomy (PHYS, small, only 15.62% of women) for the 2000 cohort tell fundamentally different stories. In BIO, women are characterized by markedly lower survival rates than men, with the divergence increasing with each successive time interval, and in PHYS, the survival rates for men and women for two decades are nearly identical. Compared with men, women in BIO disappeared from science in ever-larger proportions over two decades; in contrast, women in PHYS disappeared in almost exactly the same proportions as men for the entire period examined.

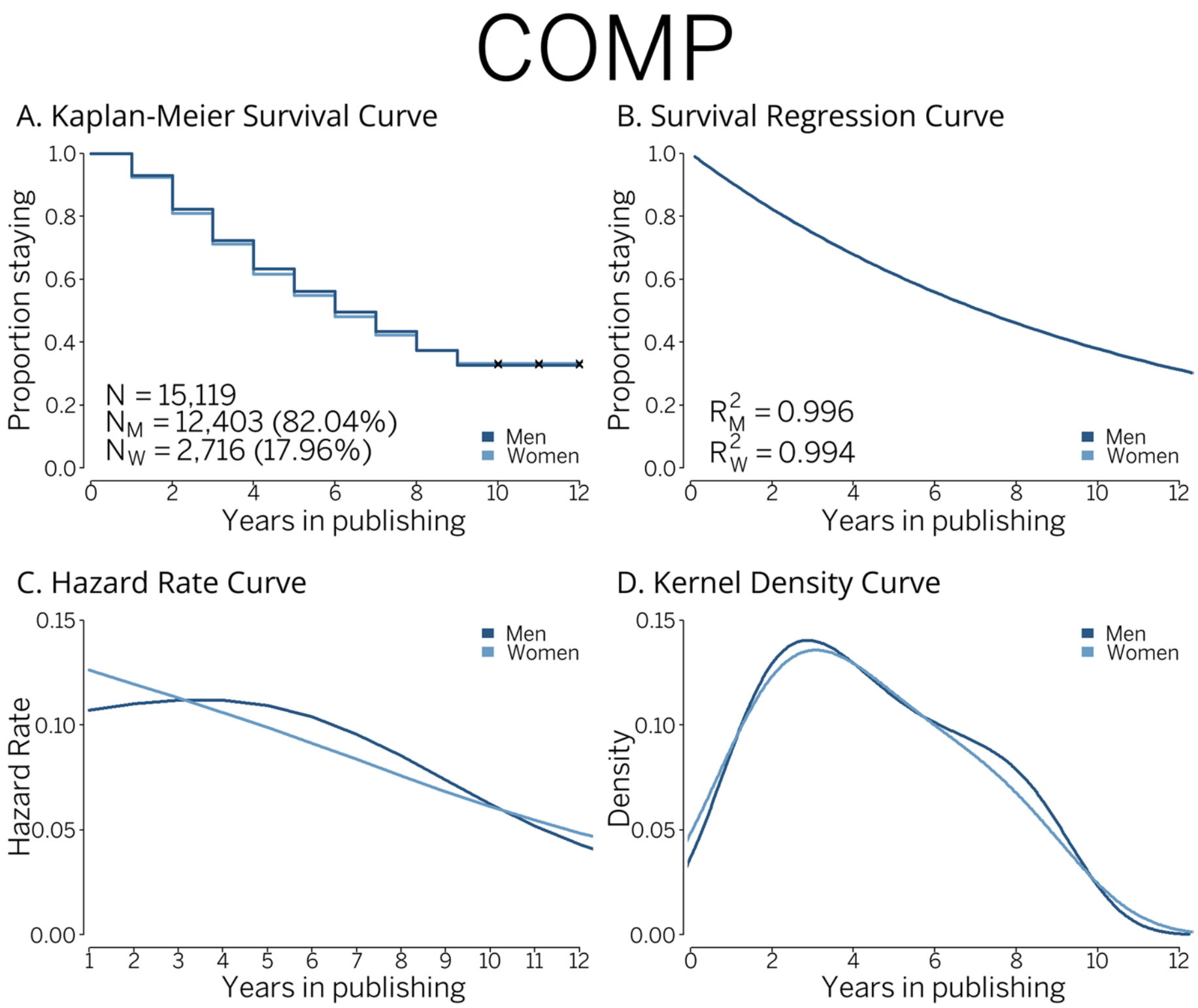

**Fig. 10** Kaplan−Meier survival curve, survival regression curve (exponential distribution fitting of Kaplan−Meier curve), hazard rate curve (B-splines smoothing method, smoothing parameter 10 k), and kernel density curve (B-splines smoothing method, bandwidth 2, component per point based on Gaussian curve), COMP, the 2010 cohort of scientists ($N=15{,}119$)

However, for the 2010 cohort, the dramatic lack of differences in attrition for men and women for all disciplines combined was much less pronounced, but gender differences in attrition still existed among disciplines. In both cases, in math-intensive disciplines, such as MATH, COMP, and PHYS, gender differences are nonexistent. In highly math intensive disciplines, the share of women entering science is small or very small, but women stay on in science in exactly the same proportions as men—which may suggest that they are extremely professionally successful, despite possible unwelcoming workplace climates. In disciplines with very low representation of women, where women are very visible minorities (constituting 20% or less of all publishing scientists and often playing the role of "tokens," or exemplary figures in university departments) (Kanter, 1977), the newcoming and surviving women stay in the system of science as powerfully as men do. In these male-dominated fields, there is a heavy gendered selection; women who succeed in publishing two research articles (our sampling selection criteria) are more likely to survive as publishing scientists in the long run than women in less math-intensive fields.

Taking computing as another example, the presence of publishing women in COMP has slowly increased over time compared with that in other STEMM disciplines, and women in COMP are rare in senior academic ranks, which is crucial to their decision-making capacities (Fox & Kline, 2016). Branch and Alegria (2016) reported gendered obstacles and "persistent exclusionary messages" to women that they do not belong to computing. Our data on COMP, as well as MATH, PHYS, and ENG, may indicate that women in these fields may not be like women academics on average in STEMM fields, especially compared with the large and heavily women-populated MED, BIO, and AGRI fields. Successful women in these disciplines may be "different," "atypical," not the average women at the center of the distribution (Branch & Alegria, 2016). Their perceived chances of promotion from associate to full professors are most heavily dependent on a departmental climate perceived to be stimulating/collegial (Fox & Xiao, 2013). The academic ranks—not merely the presence—of women in COMP are important because the ranks are where status and decision-making belong in universities (Fox & Kline, 2016).

Women have traditionally been believed (Goulden et al., 2011; Preston, 2004, 2014; Shaw & Stanton, 2012; Wolfinger et al., 2008) to be leaving science earlier and in higher proportions than men (which is generally confirmed in our analyses of the 2000 cohort), leading to much higher attrition rates across many disciplines. For more recent cohorts, in contrast, gender differences in attrition rates may no longer be present, especially for math-intensive disciplines with low numbers and percentages of women. For new generations of scientists, attrition in science has been on the rise and very high (58.6% of women and 57.6% of men from the 2010 cohort disappeared from science or ceased publishing within 9 years), but it seems to be much less gendered than traditionally assumed.

Changes in the participation of women in science over the past three decades have been tectonic, and bibliometric-based longitudinal studies of male and female scientists have opened new opportunities for global cross-gender and cross-cohort analyses. In a fast-changing science environment (Stephan, 2012; Wang & Barabási, 2021), with hundreds of thousands of newcomers to science every year, our traditional assumptions about how men and women disappear from science may need careful revision, and our intention was to sketch some tentative general answers and possible directions for further, more detailed studies.

**Supplementary Information** The online version contains supplementary material available at https://doi.org/10.1007/s10734-024-01284-0.

**Acknowledgements** Marek Kwiek would like to thank the hosts and audiences of his seminars on attrition: DZHW Berlin (Torger Möller, November 2023), the University of Hong Kong (Hugo Horta, January 2024), the University of Oxford (Simon Marginson, January 2024), Leiden University (Ludo Waltman, February 2024), Stanford University (John Ioannidis, March 2024), and the Association for Information Science & Technology (Chenwei Zhang, Hong Kong, May 2024). We gratefully acknowledge the assistance of the International Center for the Studies of Research (ICSR) Lab and Kristy James, Senior Data Scientist. We also want to thank Dr. Wojciech Roszka from the CPPS Poznan Team for many fruitful discussions.

**Funding** MK gratefully acknowledges the support provided by MEiN NDS grant no. NdS/529032/2021/2021. LS is grateful for the support of his doctoral studies provided by the NCN grant 2019/35/0/HS6/02591.

## Declarations

**Conflict of interest** The authors declare no competing interests.

## References

Allison, P. D. (2014). *Event history and survival analysis*. Sage.

Alper, J. (1993). The pipeline is leaking women all the way along. *Science, 260*(5110), 409–411.

Baas, J., Schotten, M., Plume, A., Côté, G., & Karimi, R. (2020). Scopus as a curated, high-quality bibliometric data source for academic research in quantitative science studies. *Quantitative Science Studies, 1*(1), 377–386. https://doi.org/10.1162/qss_a_00019

Blickenstaff, J. C. (2005). Women and science careers: Leaky pipeline or gender filter? *Gender and Education, 17*(4), 369–386.

Branch, E. H. (Ed.). (2016). *Pathways, potholes, and the persistence of women in science*. Lexington Books.

Branch, E. H., & Alegria, S. (2016). Gendered responses to failure in undergraduate computing. Evidence, contradictions, and new directions. In E.H. Branch (Eds.), *Pathways, potholes, and the persistence of women in science: Reconsidering the pipeline* (pp. 17–31). Lexington Books.

Britton, D. M. (2017). Beyond the chilly climate: The salience of gender in women's academic careers. *Gender & Society, 31*(1), 5–27.

Cornelius, R., Constantinople, A., & Gray, J. (1988). The chilly climate: Fact or artifact? *The Journal of Higher Education, 59*(5), 527–555.

Deutsch, F. M., & Yao, B. (2014). Gender differences in faculty attrition in the USA. *Community, Work & Family, 17*(4), 392–408.

Ehrenberg, R. G., Kasper, H., & Rees, D. I. (1991). Faculty turnover in American colleges and universities. *Economics of Education Review, 10*(2), 99–110.

Elsevier (2018). *Gender in the global research* landscape. Elsevier

Elsevier (2020). *The researcher journey through a gender lens*. Elsevier

Fox, M. F. (2010). Women and men faculty in academic science and engineering: Social-organizational indicators and implications. *American Behavioral Scientist, 53*, 997–1012.

Fox, M. F. (2020). Gender, science, and academic rank: Key issues and approaches. *Quantitative Science Studies, 1*(3), 1001–1006.

Fox, M. F., & Mohapatra, S. (2007). Social-organizational characteristics of work and publication productivity among academic scientists in doctoral-granting departments. *The Journal of Higher Education, 78*(5), 542–571.

Fox, M. F., & Xiao, W. (2013). Perceived chances for promotion among women associate professors in computing: Individual, departmental, and entrepreneurial factors. *Journal of Technology Transfer, 38*, 135–152.

Fox, M. F., Realff, M. L., Rueda, D. R., & Morn, J. (2017). International research collaboration among women engineers: Frequency and perceived barriers, by regions. *Journal of Technology Transfer, 42*(6), 1292–1306.

Fox, M.F., & Kline, K. (2016). Women faculty in computing. A key case of women in science. In: E.H. Branch (ed.), *Pathways, potholes, and the persistence of women in science. Reconsidering the pipeline* (p. 54–69). Lexington Books.

Geuna, A., & Shibayama, S. (2015). Moving out of academic research: Why do scientists stop doing research? In A. Geuna (Ed.), *Global mobility of research scientists* (pp. 271–297). Elsevier.

Glenn, N. D. (2005). *Cohort Analysis*. Sage.

Goulden, M., Mason, M. A., & Frasch, K. (2011). Keeping women in the science pipeline. *The ANNALS of the American Academy of Political and Social Science, 638*, 141–162.

Hammarfelt, B. (2017). Recognition and reward in the academy: Valuing publication oeuvres in biomedicine, economics and history. *Aslib Journal of Information Management, 69*(5), 607–623.

Holman, L., Stuart-Fox, D., & Hauser, C. E. (2018). The gender gap in science: How long until women are equally represented? *PLOS Biology, 16*(4), 1–20.

Ioannidis, J. P. A., Boyack, K. W., & Klavans, R. (2014). Estimates of the continuously publishing core in the scientific workforce. *PLoS ONE, 9*(7), e101698.

Kaminski, D., & Geisler, C. (2012). Survival analysis of faculty retention in science and engineering by gender. *Science, 335*, 864–866.

Kanter, R. M. (1977). Some effects of proportions on group life: Skewed sex ratios and responses to token women. *American Journal of Sociology, 82*(5), 965–990.

Karimi, F., Wagner, C., Lemmerich, F., Jadidi, M., & Strohmaier, M. (2016). Inferring gender from names on the web: A comparative evaluation of gender detection methods. In *Proceedings of the 25th international conference companion on world wide web* (pp. 53–54).

Kashyap, R., Rinderknecht, R. G., Akbaritabar, A., Alburez-Gutierrez, D., Gil-Clavel, S., Grow, A., Zhao, X. (2023). Digital and computational demography. In: *Research handbook on digital sociology*, 48–86. Cheltenham: Edward Elgar Publishing.

King, M. M., Bergstrom, C. T., Correll, S. J., Jacquet, J., & West, J. D. (2017). Men set their own cites high: Gender and self-citation across fields and over time. *Socius, 3*

Kwiek, M. (2016). The European research elite: A cross-national study of highly productive academics across 11 European systems. *Higher Education, 71*(3), 379–397.

Kwiek, M. (2018). High research productivity in vertically undifferentiated higher education systems: Who are the top performers? *Scientometrics, 115*(1), 415–462. https://doi.org/10.1007/s11192-018-2644-7

Kwiek, M. (2019). *Changing European academics: A comparative study of social stratification, work patterns and research productivity*. Routledge.

Kwiek, M., & Roszka, W. (2021a). Gender disparities in international research collaboration: A large-scale bibliometric study of 25,000 university professors. *Journal of Economic Surveys, 35*(5), 1344–1388.

Kwiek, M., & Roszka, W. (2021b). Gender-based homophily in research: A large-scale study of man-woman collaboration. *Journal of Informetrics, 15*(3), 1–38.

Kwiek, M., & Roszka, W. (2022). Are female scientists less inclined to publish alone? The gender solo research gap. *Scientometrics, 127*, 1697–1735.

Kwiek, M., & Roszka, W. (2023). The young and the old, the fast and the slow: A large-scale study of productivity classes and rank advancement. *Studies in Higher Education*. https://doi.org/10.1080/03075079.2023.2288172

Kwiek, M., & Roszka, W. (2024). Once highly productive, forever highly productive? Full professors' research productivity from a longitudinal perspective. *Higher Education, 87*, 519–549.

Kwiek, M., & Szymula, L. (2023). Young male and female scientists: A quantitative exploratory study of the changing demographics of the global scientific workforce. *Quantitative Science Studies, 4*(4), 902–937.

Larivière, V., Ni, C., Gingras, Y., Cronin, B., & Sugimoto, C. R. (2013). Bibliometrics: Global gender disparities in science. *Nature, 504*(7479), 211–213.

Liu, L., Jones, B. F., Uzzi, B., et al. (2023). Data, measurement, and empirical methods in the science of science. *Nature Human Behaviour, 7*, 1046–1058.

Maranto, C. L., & Griffin, A. E. (2011). The antecedents of a "chilly climate" for women faculty in higher education. *Human Relations, 64*(2), 139–159.

Menard, S. (2002). *Longitudinal Research*. Sage.

Mihaljević, H., & Santamaría, L. (2020). Authorship in top-ranked mathematical and physical journals: Role of gender on self-perceptions and bibliographic evidence. *Quantitative Science Studies, 1*(4), 1468–1492.

Mihaljević-Brandt, H., Santamaría, L., & Tullney, M. (2016). The effect of gender in the publication patterns in mathematics. *PLoS ONE, 11*(10), 1–3.

Mills, M. (2011). *Introducing survival and event history analysis*. Sage.

Milojevic, S., Radicchi, F., & Walsh, J. P. (2018). Changing demographics of scientific careers: The rise of the temporary workforce. *Proceedings of the National Academy of Sciences, 115*, 12616–12623.

NamSor (2024). NamSor API documentation available from https://namsor.app/api-documentation/

Nielsen, M. W., & Andersen, J. P. (2021). Global citation inequality is on the rise. *Proceedings of the National Academy of Sciences, 118*(7), e2012208118.

Nygaard, L. P., Piro, F., & Aksnes, D. (2022). Gendering excellence through research productivity indicators. *Gender and Education, 34*(6), 690–704.

Preston, A. E. (2004). *Leaving science*. Russell Sage Foundation.

Rosser, V. J. (2004). Faculty members' intentions to leave: A national study on their work-life and satisfaction. *Research in Higher Education, 45*(3), 285–309.

Sanliturk, E., Zagheni, E., Dańko, M. J., Theile, T., & Akbaritabar, A. (2023). Global patterns of migration of scholars with economic development. *PNAS, 120*(4), e2217937120.

Santamaría, L., & Mihaljević, H. (2018). Comparison and benchmark of name-to-gender inference services. *PeerJ Computer Science, 4*, e156. https://doi.org/10.7717/peerj-cs.156

Santos, J. M., Horta, H., & Amâncio, L. (2020). Research agendas of female and male academics: A new perspective on gender disparities in academia. *Gender and education, 33*(5), 625–43.

Science-Metrix (2018). Analytical support for bibliometrics indicators. Development of bibliometric indicators to measure women's contribution to scientific publications. Final report. Science-metrix.

Sebo, P. (2021). Performance of gender detection tools: A comparative study of name-to-gender inference services. *Journal of the Medical Library Association, 109*(3), 414–421.

Sebo, P. (2023). How well does NamSor perform in predicting the country of origin and ethnicity of individuals based on their first and last names? *PLOS One, 18*(11), e0294562. https://doi.org/10.1371/journal.pone.0294562

Shaw, A. K., & Stanton, D. E. (2012). Leaks in the pipeline: Separating demographic inertia from ongoing gender differences in academia. *Proceedings of the Royal Society B Biological sciences, 279*(1743), 3736–3741.
Shibayama, S., & Baba, Y. (2015). Impact-oriented science policies and scientific publication practices: The case of life sciences in Japan. *Research Policy, 44*(4), 936–950.
Singer, J. D. & Willett, J. B. (2003). *Applied longitudinal data analysis: Modeling change and event occurrence*. Oxford University Press.
Smart, J. C. (1990). A causal model of faculty turnover intentions. *Research in Higher Education, 31*(5), 405–424.
Spoon, K., LaBerge, N., Wapman, K. H., Zhang, S., Morgan, A. C., Galesic, M., Fosdick, B. K., Larremore, D. B., & Clauset, A. (2023). Gender and retention patterns among US faculty. *Science Advances, 9*, 2205. https://doi.org/10.1126/sciadv.adi2205
Stephan, P. E. (2012). *How economics shapes science*. Harvard University Press.
Sugimoto, C., & Larivière, V. (2023). *Equity for women in science: Dismantling systemic barriers to advancement*. Harvard University Press
Tang, L., & Horta, H. (2023). Supporting academic women's careers: Male and female academics' perspectives at a Chinese research university. *Minerva*. https://doi.org/10.1007/s11024-023-09506-y
Wang, D., & Barabási, A.-L. (2021). *The science of science*. Cambridge University Press.
Watt, H. M. G., Hyde, J. S., Petersen, J., Morris, Z. A., Rozek, C. S., & Harackiewicz, J. M. (2017). Mathematics - A critical filter for STEM-related career choices? A longitudinal examination among Australian and U.S. adolescents. *Sex Roles, 77*, 254–271.
West, J. D., Jacquet, J., King, M. M., Correll, S. J., & Bergstrom, C. T. (2013). The role of gender in scholarly authorship. *PLoS ONE, 8*(7), e66212.
White-Lewis, D. K., O'Meara, K., Mathews, K., et al. (2023). Leaving the institution or leaving the academy? Analyzing the factors that faculty weigh in actual departure decisions. *Research in Higher Education, 64*, 473–494.
Wohrer, V. (2014). To stay or to go? Narratives of early-stage sociologists about persisting in academia. *Higher Education Policy, 27*, 469–487.
Wolfinger, N. H., Mason, M. A., & Goulden, M. (2008). Problems in the pipeline: Gender, marriage, and fertility in the ivory tower. *Journal of Higher Education, 79*(4), 388–405.
Xie, Y., & Shauman, K.A. (2003). *Women in science: Career processes and outcomes*. Harvard University Press
Xu, Y. J. (2008). Gender disparity in STEM disciplines: A study of faculty attrition and turnover intentions. *Research in Higher Education, 49*, 607–624.
Zhou, Y., & Volkwein, J. F. (2004). Examining the influence on faculty departure intentions: A comparison of tenured versus nontenured faculty at research universities using NSOPF-99. *Research in Higher Education, 45*(2), 139–176.